\documentclass[journal]{IEEEtran}
\usepackage{array}
\usepackage{makecell}

\usepackage{amsmath,amssymb,amsfonts}
\usepackage{algorithmic}
\usepackage{graphicx}
\usepackage{textcomp}
\usepackage[dvipsnames]{xcolor}
\usepackage[acronym]{glossaries}

\usepackage{setspace} 
\usepackage{multirow}
\usepackage[normalem]{ulem}
\usepackage{url}
\usepackage{booktabs,siunitx}
\usepackage{lineno}
\usepackage{tablefootnote}
\usepackage{arydshln}
\usepackage{placeins}

\usepackage{amsmath}
\usepackage[linesnumbered,ruled]{algorithm2e}

\usepackage[
natbib=true,
style=numeric,
sorting=none
]{biblatex}

\usepackage{tabularray}

\newcolumntype{?}{!{\vrule width 2pt}}
\newacronym{hs}{HS}{Hyperspectral}
\newacronym{anc}{ANC}{Abundance Non-negativity Constraint}
\newacronym{asc}{ASC}{Abundance Sum-to-one Constraint}
\newacronym{eea}{EEA}{Endmember Extraction Algorithms}
\newacronym{cnn}{CNN}{Convolution Neural Network}
\newacronym{an}{PC}{Pixel Contextualizer}
\newacronym{ae}{AE}{Autoencoder}
\newacronym{lmm}{LMM}{Linear Mixture Model}
\newacronym{ap}{AP}{Abundance Predictor}
\newacronym{sp}{SP}{Signature Predictor}
\newacronym{fs}{FusionNet}{Transformer-based Endmember Fusion with Spatial Context for Hyperspectral Unmixing}
\newacronym{rmse}{RMSE}{Root Mean Square Error}
\newacronym{sad}{SAD}{Spectral Angle Distance}

\newcommand{\bands}[0]{L}
\newcommand{\mbands}[0]{$L$}
\newcommand{\nEndM}[0]{M}
\newcommand{\mnEndM}[0]{$M$}
\newcommand{\nEEA}[0]{B}
\newcommand{\mnEEA}[0]{B}
\newcommand{\nPixel}[0]{N}
\newcommand{\mnPixel}[0]{$N$}

\newcommand{\mNeighbours}[0]{$J$}

\newcommand{\Heads}[0]{H}
\newcommand{\CAPixel}[0]{\Bar{Y}}
\newcommand{\NBPixel}[0]{\mathcal{N}}
\newcommand{\APFunc}[0]{\textbf{$g_{\Psi}$}}
\newcommand{\SPFunc}[0]{\textbf{$h_{\Phi,\Omega}$}}
\newcommand{\PCFunc}[0]{\textbf{$f_{\Theta}$}}
\newcommand{\NPN}[0]{J}

\newcommand{\Eensemble}[0]{\mathcal{S}}

\newcommand{\AP}[0]{Abundances Predictor}

\def\BibTeX{{\rm B\kern-.05em{\sc i\kern-.025em b}\kern-.08em
    T\kern-.1667em\lower.7ex\hbox{E}\kern-.125emX}}

\begin{document}
\title{Transformer based Endmember Fusion with Spatial Context for Hyperspectral Unmixing}

\author{ 
R. M. K. L. Ratnayake,~\IEEEmembership{Student Member,~IEEE,}  
D. M. U. P. Sumanasekara,~\IEEEmembership{Student Member,~IEEE,} 
H. M. K. D. Wickramathilaka,~\IEEEmembership{Student Member,~IEEE,}
G. M. R. I. Godaliyadda,~\IEEEmembership{Senior Member,~IEEE,} 
M. P. B. Ekanayake,~\IEEEmembership{Senior Member,~IEEE,} 
H. M. V. R. Herath,~\IEEEmembership{Senior Member,~IEEE,}

\thanks{All the authors of this paper are with the Department of  Electrical and Electronic Engineering, University of Peradeniya, Sri Lanka. (Email addresses are listed according to the authors' order. e-mail: e17290@eng.pdn.ac.lk, e19391@eng.pdn.ac.lk, kavindu@eng.pdn.ac.lk, roshangodd@ee.pdn.ac.lk, vijitha@eng.pdn.ac.lk, mpb.ekanayake@ee.pdn.ac.lk)}}

\markboth{Transformer based Endmember Fusion with Spatial Context for Hyperspectral Unmixing}%
{Shell \MakeLowercase{\textit{et al.}}: Bare Demo of IEEEtran.cls for Journals}

\maketitle

\begin{abstract}
In recent years, transformer-based deep learning networks have gained popularity in Hyperspectral (HS) unmixing applications due to their superior performance. The attention mechanism within transformers facilitates input-dependent weighting and enhances contextual awareness during training. Drawing inspiration from this, we propose a novel attention-based Hyperspectral Unmixing algorithm called \acrfull{fs}. This network leverages an ensemble of endmembers for initial guidance, effectively addressing the issue of relying on a single initialization. This approach helps avoid suboptimal results that many algorithms encounter due to their dependence on a singular starting point. The \acrshort{fs} incorporates a \acrfull{an}, introducing contextual awareness into abundance prediction by considering neighborhood pixels. Unlike Convolutional Neural Networks (CNNs) and traditional Transformer-based approaches, which are constrained by specific kernel or window shapes, the Fusion network offers flexibility in choosing any arbitrary configuration of the neighborhood. We conducted a comparative analysis between the \acrshort{fs} algorithm and eight state-of-the-art algorithms using three widely recognized real datasets and one synthetic dataset. The results demonstrate that \acrshort{fs} offers competitive performance compared to the other algorithms.
\end{abstract}


\IEEEpeerreviewmaketitle

\section{Introduction}

In the remote sensing field, \acrfull{hs} imaging plays a vital role. As \acrshort{hs} images contain numerous bands, they contain a lot of information compared to conventional RGB images and Multispectral images \cite{KL_Spectra_Unmixing}. Due to this, \acrshort{hs} images have become useful in a vast range of applications such as Agriculture \cite{agri_app_1, agri_app_2}, Mineral Mapping \cite{KL_rock_soil, KL_yasiru_litho}, Military applications \cite{military_app_1, military_app_2}, Environmental Monitoring \cite{KL_Earth_Observation, trends_in_hs} etc. Each pixel in the \acrshort{hs} image can be considered as a mixture of pure materials. These pure materials are called endmembers and their spectral signatures are known as endmember signatures. The percentage of each pure material that appears in a pixel is called endmember abundance. The process of extracting the spectral signature and the abundance of each endmember is known as \acrshort{hs} unmixing.

Although the spectral resolution of \acrshort{hs} images is high, they lack spatial resolution \cite{resolution_1, resolution_2}. In addition to that, \acrshort{hs} images contain noise \cite{noise_1}, \cite{noise_2}. These are the key challenges that must be addressed in \acrshort{hs} unmixing. Due to the physical meaning of the endmember abundances and signatures, three constraints should be considered in the \acrshort{hs} Unmixing process. The first constraint is that the values of the endmember signatures cannot be negative. The second constraint is the values of the endmember abundances cannot be negative and it is called \acrfull{anc}. The third one is the sum of the abundances of all the endmembers in each pixel must be one because every single pixel is a combination of these endmembers and it is called \acrfull{asc}.

Every pixel’s signature is a mixture of these endmember signatures. This mixing process can happen macroscopically and microscopically \cite{hs_overview_lmm, KL_LMM}. To model that, the \acrfull{lmm} and several non-linear mixture models have been proposed \cite{non_linear_mixing}.

Many algorithms have been proposed to solve the unmixing problem and they can be divided into four main categories. They are Geometrical methods \cite{nfindr, vca, Geometrical_nasa}, Statistical methods \cite{mewan_kbnmf, ica_based_hsu, bathiya_graph_based, KH_1, KH_2, KH_3}, Sparse Regression methods \cite{sparse1, sparse2, sparse3}, and Deep Learning methods \cite{3D_conv, cascade_ae, deep_ae_global_smooth, transformer}. Geometrical methods use either pure pixels \cite{vca, nfindr}, or minimum volume simplex to unmix the \acrshort{hs} image\cite{minvol_original, minvol_tgrs}, Statistical approach is based on the statistical properties of the \acrshort{hs} image. It tends to solve the matrix using blind source separation \cite{ICA}. When the data is highly mixed, the statistical method gives better results compared to geometrical methods as there are not enough spectral vectors in the simplex facets. The sparse-regression methods use pre-obtained laboratory signatures to construct the pixel signatures \cite{sparse_library}. In this method, the optimal subset of material signatures, that can span the pixel spectrum space is determined. Deep learning approaches use neural networks to unmix the \acrshort{hs} images. This is done either semi-supervised \cite{self_supervised} manner or unsupervised \cite{tanet_unsupervised} manner.  

Due to increasing computational power and the ability to find complex relationships, the deep learning approaches show promising results on \acrshort{hs} unmixing. A common architecture that has been used for unmixing is the Auto-encoder \cite{ae}, which consists of one encoder to compress the data to a lower dimension space (In \acrshort{hs} Unmixing, this lower dimension space can be used as abundances by setting constraints) and a decoder to reconstruct the data from the lower dimensional space. Dense layers \cite{ae, deep_ae_global_smooth}, convolutional layers \cite{cycu, 3D_conv}, and recurrent layers \cite{rnn} are being used to perform the unmixing process effectively. Many architectures that are inspired by auto-encoder, have been proposed with competitive results \cite{ae, ae_comparison, 3D_conv, cascade_ae, deep_ae_global_smooth}.

In addition to that, the recent advancements in transformer architecture \cite{attention_is_allu_need} have encouraged the application of transformer-based algorithms for \acrshort{hs} unmixing. \cite{transformer} uses a transformer encoder to unmixing \acrshort{hs} image by capturing long-range dependencies. \cite{swin_transformer} uses a swin-transformer to extract the spatial information and a simplified attention mechanism to obtain the spectral information. Using window multi-head self-attention and shifted window multi-head self-attention mechanisms, \cite{swin_transformer} can extract global spatial priors. Addressing the sparse unmixing problem, \cite{window_trans} has used a special attention mechanism named window-based pixel level multi-head self-attention. \cite{undat_double_aware_trans} exploits two properties of \acrshort{hs} images which are, region homogeneity and spectral correlation. Two modules namely, the Score-based homogeneous aware module and the spectral group aware module have been used to address these properties. \cite{spatial_spectral_attent} uses a bilateral global attention network to fuse the spectral and spatial information optimally. This architecture has two separate branches to extract the spatial and spectral features.  
\subsection{Motivation}

\label{motivation}

Deep learning methods often use endmember extraction algorithms (EEAs) \cite{vca, nfindr} to initialize the decoder part of the autoencoder. Typically, these methods have employed a single EEA for this initialization. However, the performance of EEAs has been found to vary depending on the dataset \cite{EEAS_1, impact_of_eeas}. Each EEA operates under different assumptions about the environment, and the validity of these assumptions can differ across various environments. Consequently, the EEA that performs best can vary depending on the specific environment. Moreover, the choice of candidate endmembers used to initialize an unmixing algorithm can significantly impact the results \cite{impact_of_eeas, Endmemb_init_for_NMF}. If only a single EEA is used for initialization, it may not provide the optimal starting point for a given environment, potentially leading to suboptimal results and slower convergence rates. {To elaborate, only one EEA is used to initialization of an algorithm, there is a high potential for the algorithm to be bound by the assumptions and constraints governing that EEA. This can hinder performance and convergence.} Despite this observation, there is a notable lack of research exploring the use of an ensemble of signatures for each endmember, obtained from different EEAs, for purposes of initial guidance of the network.




In hyperspectral (HS) images, as in other natural images, there is a high correlation between the pixels and their neighboring pixels. Therefore, incorporating spatial context into the HS unmixing process is crucial \cite{UST}. 
This is especially important when dealing with noisy pixels, as the context provided by neighboring pixels can mitigate the effect of noise and thereby improve accuracy. Additionally, utilizing spatial context can enhance spatial smoothness when predicting abundance maps \cite{gauss_paper}. The spatial context is often incorporated using \acrshort{cnn}s or transformers. The \acrshort{cnn}-based methods incorporate spatial context in the form of spatial correlation \cite{3D_conv, cycu}. Therefore, they only consider local pixels that are limited by the size of the kernel of the convolution layer. On the other hand, transformer-based algorithms often strive to feed the spatial context of a fixed shape to the unmixing process \cite{UST},
There is a lack of research on methods to incorporate neighborhood information of an arbitrary shape conditioned on the pixel for abundance prediction.


\subsection{Methodology Overview and Contribution}

To address the identified research gap, we propose a novel architecture called \acrshort{fs}, which consists of three key components that work in unison to predict abundance maps and endmembers for a given \acrshort{hs} image. This model integrates spatial context from various neighboring pixels through a cross-attention mechanism which mitigates the problems including spatial context in section \ref{motivation}. Additionally, the proposed model can utilize an ensemble of endmember signatures for its initial guidance, mitigating the reliance on a single initialization scheme. By starting with this ensemble, the model trains and intelligently combines the endmember signatures using the attention mechanism, ultimately producing the final endmember signatures. This addresses the main problem of over-dependence on a single initialization and related issues mentioned previously, {giving the model more flexibility to converge towards endmembers more applicable to the given environment by incorporating information from the entire ensemble.} Specifically, the main contributions of this article can be summarized as follows.  


\begin{enumerate}


    \item {We propose an architecture featuring an attention-based mechanism designed to address the challenge of unmixing performance variability dependent on environmental conditions. Our approach incorporates a learnable ensemble of endmembers, extracted from various endmember extraction algorithms, to enhance performance robustness. Notably, our network can be trained in an unsupervised manner, offering substantial performance improvements across diverse environmental conditions.}

    \item {Instead of directly applying the attention mechanism, we harness its functionalities in two distinct ways to fulfill our goals. Firstly, we employ the attention mechanism to enhance the accuracy of the predicted endmember by enabling the learnability of the ensemble of endmembers. By strategically selecting the inputs to the attention mechanism (Query, Key, and Value), we achieve superior endmember prediction compared to a simplistic linear combination. Secondly, we leverage the attention mechanism to integrate neighborhood information. We designate Query and Value as neighboring pixels, Key as the observed pixel and this strategical selection significantly improves the accuracy of pixel-wise abundance predictions while mitigating the effects of noise.}

    \item {We conduct a comprehensive ablation study, meticulously assessing how our architecture performs under diverse configurations. In particular, the effect of running with and without the \acrfull{an}, The effect of different training stages, the effect of different neighborhood shapes and sizes, the consequences of varying the number of \acrshort{eea} used to form the ensemble, the effect of noise and the effect on the number of endmembers.}

    

        

    
\end{enumerate} 

The remaining sections of the paper are as follows. Section \ref{Related Work} describes the theories and related work upon which the proposed methods builds while Section \ref{Proposed_Method} expounds upon the proposed architecture and Section \ref{experiments} describes the experimental setup and results for four different datasets and different network architectures. Finally, the concluding remarks and the summary of the findings are encapsulated in Section \ref{conclusion}

\section{Related Work}
\label{Related Work}

\subsection{Notation}
\begin{table}[h!]
\centering
\begin{tabular}{ll}
\mnEEA & Number of Endmember Extraction Algorithms \\
$N$ & Number of Pixels. \\
$L$ & Number of Spectral Bands\\
$M$ & Number of Endmembers\\
$D_H$ & Height of the \acrfull{hs} image\\
$D_W$ & Width of the \acrfull{hs} image\\
$Q$ & Query for the Attention Function\\
$K$ & Key for the Attention Function\\
$V$ & Value for the Attention Function\\
$H$ & Number of Heads in Multi-head Attention\\
$\NPN$ & Number of neighborhood pixels\\
$d_k$ & Dimension of $Q$ (and $K$) \\
$d_v$ & Dimension of $V$\\
$W_{i}^X$ & Projection Matrix of $X \in \{Q,K,V\}$\\
$W^O$ & Output Projection Matrix for Multi-Head Attention\\
$Y$ & \acrfull{hs} Image Matrix\\
$S$ & Endmember Signature Matrix\\
$A$ & Abundance Matrix\\
$E$ & Noise Matrix in \acrfull{lmm}\\
$\widehat{S}$ & Predicted Endmember Signature Matrix\\
$\widehat{A}$ & Predicted Abundance Matrix\\
$\CAPixel$ & Context Aware Pixel\\
${\Eensemble}_i$ & Ensemble of ${i^{th}}$ endmember \\ 
$\NBPixel$  & Neighbor Pixels\\
\end{tabular}
\end{table}


\subsection{Attention Function}
\label{attention_function}

\begin{figure*}
    \centering
    \includegraphics[width=0.75\textwidth]{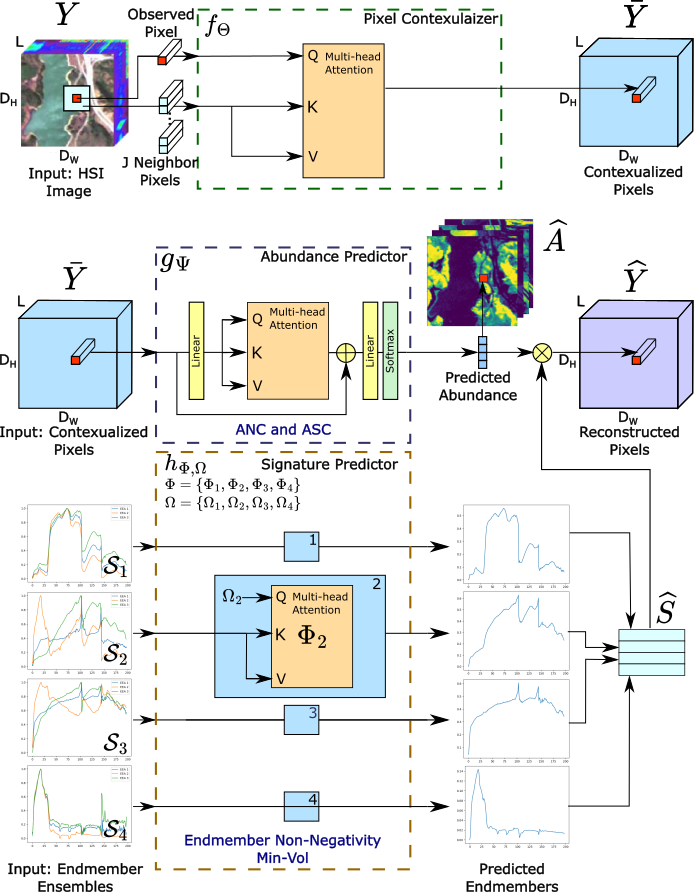}
    \caption{Graphical Representation of the Proposed \acrfull{fs} Model Illustrated for the Case of Four Endmembers}
    \label{Fusion_Architecture}
\end{figure*}

The attention mechanism introduced in \cite{attention_is_allu_need} (Scaled Dot-product attention), has demonstrated remarkable performance in a multitude of areas. As defined in the above paper, an attention function is mapping a Query ($Q$) and a set of Key-Value ($K$ and $V$) pairs to an output, where the query, keys, values, and output are all vectors. The output is computed as a weighted sum of the value parameters, where the weight assigned to each value parameter is computed by a compatibility function of the query with the corresponding key.

Due to its inherent characteristics, the model can dynamically adjust the importance it assigns to different parts of the input based on the content. Unlike linear layers, which use fixed weights, this dynamic adjustment allows the model to focus on what is relevant for specific tasks or contexts. This capability enhances the model's performance by ensuring it pays attention to the most important aspects of the input for each particular situation.

The attention function should be provided with 3 parameters as inputs namely, query $Q$, key $K$, and value $V$.  

\begin{equation} 
\text{Attention}(Q,K,V) = \text{softmax}\left(\frac{QK^T}{\sqrt{d_k}}\right)V
 \end{equation}

Considering the above formula, $QK^T$ is calculating the dot product between the query $Q$, and the keys $K$. This is then used to calculate the weights for the weighted sum as mentioned earlier using $\text{softmax}\left({\frac{QK^T}{\sqrt{d_k}}}\right)$. These weights are known as the attention weights. Multiplying the resulting weight vector by the matrix containing the values ($V) $ results in a weighted sum of the values. In this formula, $d_k$ is the dimension of the $K$ vectors (as well as $Q$). Division by $\sqrt{d_k}$ is important to avoid pushing the softmax function into regions where it has extremely small gradients as described in \cite{attention_is_allu_need}.

\subsection{Mutli-Head Attention}
\label{multihead_attention}

The Multi-Head Attention \cite{attention_is_allu_need} improves upon the Attention function by projecting the queries, keys, and values by $H$ different learnable linear layers ($H$ different 'heads') and then passing each of them to the attention function in parallel. Afterward, they are concatenated and once again projected to produce the final output.

According to \cite{attention_is_allu_need} Multi-head attention allows the model to jointly attend to information from different representation subspaces at different positions.

\begin{align}
\label{eq_multihead_attention}
\text{MultiHead}(Q,K,V) &= \text{Concat}\left(\text{head}_1, ...,\text{head}_H\right)W^O\\
\text{where } \text{head}_i &= \text{Attention}(Q{W_i}^Q,K{W_i}^K,V{W_i}^V)
\end{align} 

Where the projection parameters are ${{W_i}^Q}\in \mathbb{R}^{d_{model} \times d_k}$, ${{W_i}^K}\in \mathbb{R}^{d_{model} \times d_k}$, ${{W_i}^V}\in \mathbb{R}^{d_{model} \times d_v}$ and the output projection matrix for Multi-Head Attention is ${{W_i}^O}\in \mathbb{R}^{Hd_{v} \times d_{model}}$. Here $d_v$ is the dimension of the $V$ vectors.

In certain parts of our model, we utilize self-attention, where the three inputs to the Multi-Head Attention mechanism $Q$,$K$, and $V$ are identical. Additionally, we employ regular attention in other parts of our model, where $Q$, $K$, and  $V$ are not necessarily the same.

\subsection{\acrfull{lmm}}

In \acrshort{hs} unmixing, the \acrfull{lmm} is a widely used mixing model. Where each pixel of the image is assumed to be a linear combination of endmember signatures.  This can be mathematically represented as follows,

\begin{equation}
Y = S A + E
\end{equation}

where, $Y\in \mathbb{R}^{\bands \times \nPixel}$ is the observed \acrshort{hs} image containing  \mbands\ bands and \mnPixel\ pixels. $S\in \mathbb{R}^{\bands \times \nEndM}$ is the endmember spectral signature matrix containing \mnEndM\ endmembers. $A\in \mathbb{R}^{\nEndM \times \nPixel}$ is the abundances matrix. Here, $E\in \mathbb{R}^{\bands \times \nPixel}$ is the noise matrtix.

In addition to that, three physical constraints need to be satisfied in the unmixing problem. More precisely, the endmember matrix should be non-negative. In addition to that the abundance vectors $\mathbf{a_j}$ should satisfy the \acrfull{anc} and the \acrfull{asc} by the following equations:

\begin{equation}
\mathbf{a_j} \geq 0  \;\;\;\;\;\; \forall\; 1\leq j \leq \nEndM
\end{equation}

\begin{equation}
\sum\limits_{j=1}^\nEndM \mathbf{(a_j)} = 1
\end{equation}
\subsection{\acrlong{ae}}

\acrfull{ae} is a type of artificial neural network designed for unsupervised learning that learns low-dimensional representations of data by encoding and reconstructing input patterns. Comprising an encoder and a decoder, it aims to capture essential features, reducing dimensionality while maintaining the crucial information needed for accurate reconstruction.

Encoder: The encoder transforms the ${\mathbf{i}^{ th}}$ input vector $X_{i}\in \mathbb{R}^{\bands}$ into a hidden representation $T_i\in \mathbb{R}^{\nEndM}$ by utilizing some
trainable network parameters. In general, this can be expressed as,

\begin{equation}
T_i = f_E (X_i)
\end{equation}

where ${f_E}$ denotes the encoder transformation.

Decoder: In an \acrshort{ae} architecture, the decoder is responsible for reconstructing the original data from the low-dimensional representation produced by the encoder network. Usually, in the context of \acrshort{hs} Unmixing, the decoder is responsible for reconstructing the original pixel from the abundance values produced by the encoder network.

\begin{equation}
\widehat{X_i} = f_D (T_i)
\end{equation}
In our proposed method, a modified \acrlong{ae} architecture is used.

\section{Proposed Method}
\label{Proposed_Method}

\begin{table*}[t]
\centering
\caption{The Proposed Architecture's Layer Description}
\footnotesize
\begin{tabular}{c |c |c |c |c |c |c |c}
\Xhline{2pt}
\multirow{2}{4em}{Network} & \multirow{2}{4em}{Layers} & \multicolumn{6}{c}{Dimensions}\\
\cline{3-8}
& & Number of heads & Q & K & V & Input & Output\\
\Xhline{2pt}
\acrlong{an} & Multihead Attention & $\Heads_{\acrshort{an}}$ & $\bands \times \mNeighbours$ & $\bands \times 1$ & $\bands \times \mNeighbours$ & N/A & $\bands \times 1$\\
\hline
\multirow{3}{4em}{\acrlong{ap}} & Linear Layer & N/A & N/A & N/A & N/A & $\bands \times 1$ & $\bands \times 1$\\
\cline{2-8}
& Multihead Attention & $\Heads_{\acrshort{ap}}$ & $\Heads_{\acrshort{ap}} \times L$ & $\Heads_{\acrshort{ap}} \times L$ & $\Heads_{\acrshort{ap}} \times L$ & N/A & $\Heads_{\acrshort{ap}} \times L$ \\
\cline{2-8}
& Linear Layer & N/A & N/A & N/A & N/A & $\bands \times 1$ & $\nEndM \times 1$\\
\hline
\acrlong{sp} & MutliHead Attention $\times$ \mnEndM & $H_{\acrshort{sp}}$ & $\bands \times \nEEA$ & $\bands \times 1$ & $\bands \times \nEEA$ & N/A & $\bands \times 1$\\
\Xhline{2pt}
\end{tabular}
\label{table: network parameters}
\end{table*}


The proposed architecture consists of three main components, namely \acrfull{an}, \acrfull{ap}, and \acrfull{sp}. The architecture is shown in Figure \ref{Fusion_Architecture}.  

The \acrshort{an} generates a context-aware pixel for each pixel by using it and its neighboring pixels as inputs.
The purpose of \acrshort{ap} is to take each context-aware pixel as input and predict the corresponding abundance values of the considered pixel. \acrshort{sp} predicts the spectral signatures of the endmembers using the provided endmember ensembles as an initial starting point (In this paper the ensembles are obtained via \acrshort{eea}s). As mentioned previously the model is capable of updating and training the ensemble itself in addition to intelligently combining them using the attention mechanism to predict the endmemeber signature. The exact mechanism will be discussed in later sections.

The \acrshort{ap} and \acrshort{sp} are trained simultaneously. Where during training, the \acrshort{hs} image is reconstructed ($\widehat{Y}$) by taking the matrix product between the predicted abundances ($\widehat{A}$) and the predicted endmember signatures ($\widehat{S}$). By optimizing for the loss between the reconstructed and original images along with several regularizing terms the model converges to the correct abundance and endmember values.


A layer-wise description of the architecture is summarized in table \ref{table: network parameters}. In the following sections, these components are discussed in detail.
\subsection{\acrfull{an}}
\label{sec:PC}
The \acrfull{an} is for generating pixels that consist of spatial context. This is done as a pixel-wise operation by taking each pixel ($y_k$) in the \acrshort{hs} image and its neighborhood ($\eta_k$) as the input and giving the context-aware pixel ($\Bar{y}_k$) as the output. 


Architecture of the \acrshort{an} consist of a multi-head attention block with $H_{\acrshort{an}}$ heads. A given pixel ($y_k \in \mathbb{R}^L$) is provided as the Query for the block. The pixel's ($y_k$'s) ( Neighbourhood ($\eta_k \in \mathbb{R}^{\NPN \times L} $) where $\NPN$ is the number of neighbor pixels, is provided as the Keys and Values. As explained in \ref{attention_function} and \ref{multihead_attention} sections, the calculation of the dot product between the Query (Q) and the Key (K) which in this case is the Pixel and its Neighbourhood, calculates the correlation between the Pixels and each of its Neighbouring Pixels. These are then used as weights to combine the Neighbourhood Pixels (the Value parameter ($V$) to produce the output of the \acrshort{an}.

The presence of the linear layers in multiple head configuration, allows the model to attend to different subspace representations at different parts of the signature to produce a more optimal output. Since the output is produced by combining the Neighbourhood Pixels of a given pixel guided by the pixel itself, the produced pixel is similar to the original pixel but contains the contextual information of the neighborhood which also improves spatial smoothness.

\begin{figure}[!t]
    \includegraphics[width=\linewidth]{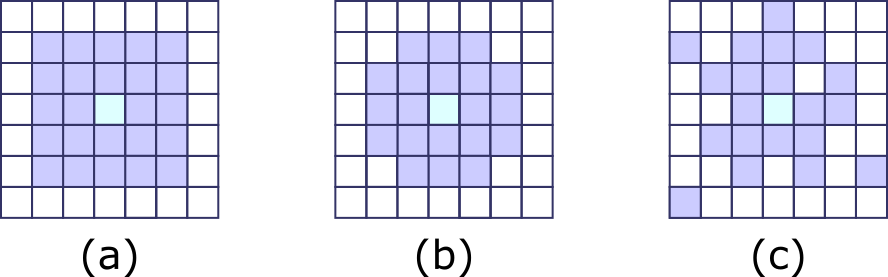}%

  \caption{Different Neighborhood Selecting Criteria.(a) Doughnut, (b) Circle, (c) Random Normal}
  \label{neighborhood_configurations}
\end{figure}

When considering how neighbors are selected in the proposed \acrshort{fs} algorithm, there is flexibility to select neighbors in any configuration in contrast to kernels used in CNNs. For this paper, three configurations have been explored. These are Doughnut configuration, Circle configuration, and Random Normal configuration (Randomly sampled from a normal distribution). Their shapes are shown in Figure \ref{neighborhood_configurations}.


As a summary, the function done by \acrshort{ap} is given in equation \ref{pc_func}

\begin{equation}
{\CAPixel} = \PCFunc(Y,\NBPixel)
\label{pc_func}
\end{equation}

Where $\Theta$ represent the trainable parameters of the model.

\subsection{\acrfull{ap}}

The task of this is to predict abundance values. This is done in a pixel-wise manner taking each of the context-aware pixels (which is produced by the \acrshort{an}) as inputs and predicting the abundance values of each of the pixels.


The \acrshort{ap} consists of four main blocks, a linear layer, a multi-head attention layer, a linear layer, and finally a softmax layer as can be seen from Figure \ref{Fusion_Architecture}. As illustrated in the figure, the model also includes a residual connection to combat the vanishing gradient problem and improve convergence. The model employs self-attention (the same input is given to $Q$,$K$ and $V$) using a multi-head attention block with $H_{\acrshort{ap}}$ heads for the purpose of predicting abundance values. In this case, the purpose of the multi-head attention layer is simply as a transforming layer that depends on the input and is thus more general than a simple linear layer. The final softmax layer enforces the \acrshort{anc} and \acrshort{asc} constraints on the predicted abundances.

The goal of the \acrshort{ap} is to predict pixel-by-pixel abundances using the spatial context learned by the \acrshort{an}. These predicted abundances, along with the predicted endmembers from the Signature Predictor (SP), aim to accurately reconstruct the original hyperspectral (HS) image. The \acrshort{an} generates contextualized pixels by leveraging Neighborhood Pixel Correlation, injecting context-awareness into the process. This context is then used by the AP for its pixel-by-pixel abundance prediction. It is important to note that the reconstruction loss is calculated in a pixelwise manner against the original HS image pixels. Reducing the reconstruction loss ensures the convergence of the outputs to the correct abundance values.

As a summary, the function done by \acrshort{ap} is given in equation \ref{ap_func}

\begin{equation}
\widehat{A} = \APFunc(\CAPixel)
\label{ap_func}
\end{equation}

Where $\Psi$ represents the trainable parameters of the model.

\subsection{\acrfull{sp}}
\label{sp_architecture}
The \acrlong{sp} takes an ensemble of possible signatures for each endemember ($\{\Eensemble_i\}_{i=1}^M$) and predicts endmember signatures for the dataset. These ensembles are obtainable using different methods. For this study, the initial ensembles are obtained via \acrshort{eea}s.

To further elaborate on the model's mechanisms, the model is designed to update and modify the ensemble itself, ensuring that it accurately represents the true endmembers. By utilizing the cross-attention mechanism, the model intelligently selects optimal weights based on the current ensemble. These weights, which are dependent on the ensemble at that specific moment, are then used to combine the ensemble. This combination results in the predicted endmember signatures. This approach mitigates the over-reliance on a single initialization, which could otherwise cause the model to converge to a local minimum dictated by the initial setup. Guided by multiple potential signatures that evolve during the training process, the model has more options and can converge to the signatures most representative of the environment. It is important to note that since the endmember ensemble updates throughout the training process, the final ensemble will differ from the initial ensemble obtained via \acrshort{eea}s. The initial ensemble serves merely as a starting point, allowing the ensemble to evolve, ensuring that the final endmember predictions are not a linear combination of the \acrshort{eea}s.

The \acrshort{sp} contains several multi-head attention blocks (with $H_{\acrshort{sp}}$ heads) equal to the number of endmembers (\(\nEndM\)) in the dataset. Each block is dedicated to predicting a single endmember signature. The \(i^{th}\) block operates on the \(i^{th}\) ensemble (\(\Eensemble_i\)), which serves as the starting ensemble for the \(i^{th}\) endmember. Each ensemble comprises \(\mnEEA\) initial candidate signatures for each endmember signature. See Figure \ref{Fusion_Architecture}.

The $i^{th}$ endmember ensemble is provided as the Key ($K$) and Value ($V$) for the $i^{th}$ Multi-head Attention Block. For the Query ($Q$), a trainable parameter vector $\Omega_i \in \mathbb{R}^{L}$ is used. The attention weights are calculated using the trainable query vector ($\Omega_i$) and the transformed ensemble, as detailed in section \ref{attention_function}. This mechanism enables the model to compute attention weights that are conditioned on the ensemble itself. Additionally, the Linear Layer inside the Multi-head Attention Block corresponding to the Value parameter ($W_V$) can transform the incoming ensemble. This enables the model to fine-tune the ensemble before it is combined using the calculated weights. In other words, the model updates and trains the ensemble before combining it with the calculated attention weights to produce the final predicted endmember. The final output projection layer ($W^O$) of the Multi-head Attention Block further transforms the predicted endmember, providing the model with additional flexibility.

The set of trainable query parameters for each of the $M$ Multi-head Attention blocks is denoted as $\Omega = \{\Omega_i\}_{i=1}^M$. The trainable parameters within the $i^{th}$ Multi-head Attention block, specifically the weights of the linear layers, are represented by $\Phi_i$. Collectively, the set of these parameters for all $M$ blocks is denoted as $\Phi = \{\Phi_i\}_{i=1}^M$.

The function performed by \acrshort{sp} can be summarized in equation \ref{sp_func},

\begin{equation}
\widehat{S} = \SPFunc(\{\Eensemble_i\}_{i=1}^M)\\
\label{sp_func}
\end{equation}

To ensure convergence, we methodically control the training of the parameters \(\Phi\) and \(\Omega\). Additionally, we incorporate regularizing terms into the training process. These aspects will be discussed in detail in a later section.

\subsection{Pixel Reconstruction}

Since \acrshort{lmm} is used for this unmixing process, the matrix product between the predicted abundances ${\widehat{A}}$ by the \AP \ and the predicted endmember signatures ${\widehat{S}}$ (whose $i^{\text{th}}$ column is the $i^{\text{th}}$ predicted endmember $\widehat{S}_i$) by the \acrshort{sp} gives the reconstructed pixel ${\widehat{Y}}$.

\begin{equation}
 {{\widehat{Y}} = \widehat{S} \; \widehat{A} }   
\end{equation}
\subsection{Constrains}

Before discussing how each of the constraints was used in the different stages, a brief explanation of the mathematical formulation and the conceptual interpretation of the constraints are presented.

\subsubsection{Mean Square Error (MSE)}

Mean Square Error (MSE) is the square of the Euclidean distance of two vectors as given in the equation \ref{L_MSE}

\begin{equation}
\mathcal{L}_{MSE}(\mathbf{\widehat{y}},\mathbf{y}) = \frac{1}{N}\sum\limits_{i=1}^N (\widehat{y_i} - y_i)^2 
\label{L_MSE}
\end{equation}

\subsubsection{Spectral Angle Distance (SAD)}

Spectral Angle Distance is the angular distance between two vectors and it is calculated as given in the equation \ref{L_SAD}
\begin{align}
    \mathcal{L}_{SAD}(\mathbf{\widehat{y}},\mathbf{y}) = \arccos \left({\frac{\mathbf{\widehat{y}^T} \mathbf{y}}{||\mathbf{\widehat{y}}||\;||\mathbf{y}||}}\right)
\label{L_SAD}   
\end{align}

where $||\cdot||$ reprensts the 2-norm of the vector.
\subsubsection{Minimum Volume Constrain (Min-Vol)}
\label{sec:min_vol_calc}
Guidance by \acrshort{asc} and \acrshort{anc} constraints is not enough to guarantee proper convergence of the \acrshort{sp} to the accurate end member signatures. This is because there is a large solution space that satisfies the above constraints. The desired solution in the solution space is the endmember signatures that form the minimum volume simplex \cite{minvol_jstar}, \cite{minvol_original}, \cite{minvol_tgrs}

To address this problem, we constrain the volume of the simplex formed by the endmembers, to be less than a certain threshold (How this threshold is determined will be explained later). To calculate the simplex volume, the method proposed by the N-FINDR paper \cite{nfindr} was used. The mathematical formulation of the method is presented below.

\begin{align}
    S_{\text{proj}} &= \begin{bmatrix}
        1 & 1 & \dots &1\\
        s_1 & s_2 & \dots & s_M
    \end{bmatrix}\\
    \mathcal{V}({S}) &= \frac{1}{(M-1)!}\text{abs}(|{S_{\text{proj}}}|)
\label{minVol}
\end{align}

$|\cdot|$ represents the determinant of the matrix. $S_{\text{proj}} \in \mathbf{R}^{M\times M}$ is formed by $\{s_{\text{proj}_i}\}_{i=1}^M$ which are the endmembers projected to a $(M-1)$ dimensional space. PCA fitted to the pixel space was used to calculate the projection matrix used for this dimensionality reduction.

The loss is calculated by taking the ReLU (Rectified Linear Unit) of the difference between the threshold value and the calculated volume of the simplex.

\begin{align}
    \mathcal{L}_{Min-Vol}(S,V_T) &= ReLU\left(\mathcal{V}(S) - V_T\right)
\end{align}

Where,
\begin{equation}
\text{ReLU}(x) = \begin{cases} 
0 & \text{if } x \leq 0 \\
x & \text{if } x > 0 
\end{cases}
\end{equation}

\begin{algorithm}
    \SetKwInOut{Input}{Input}
    \SetKwInOut{Output}{Output}

    \Input{\\
    Observed Pixels: $Y = \{y_i\}_{i=1}^N$,\\
    Neighborhood shape,\\
    Neighborhood size,\\
    Learning rate: $\alpha$,\\
    Batch Size : $D$,\\
    Epochs\\}
    \Output{\\
    Context-Aware Pixels: $\CAPixel$\\
    }
    \BlankLine
    Initialize Model Parameters $\Theta$;\\
    Neighbour Pixels $\NBPixel$: $\{{\eta_k}\}_{k=1}^N$;\\

    \For{$\textnormal{k}=1$ \KwTo ${N}$}{
        Based on Neighborhood shape and Neighborhood size,\\
        $\eta_k \gets \textnormal{Neighbour Pixels of pixel k ($y_k$)}$;\\
    }
    \BlankLine
    
    \textbf{Training Stage:}\\
    
    \For{$\textnormal{epoch}=1$ \KwTo $\textnormal{Epochs}$}{
        \For{$\textnormal{batch}=1$ \KwTo $\left\lceil \frac{N}{D}\right\rceil$}{
            Sample $D$ pixels ($Y_{\textnormal{batch}}$) from $(Y)$;\\
            Get Corresponding Neighbors (${\NBPixel}_{\textnormal{batch}}$) from $\NBPixel$\\
            \acrlong{an}: \PCFunc;\\
            $\CAPixel_\textnormal{batch}$ = $\PCFunc(Y_\textnormal{batch},{\NBPixel}_{\textnormal{batch}})$;\\

            Loss : $\mathcal{L} = \mathcal{L}_{MSE}(\CAPixel,Y)$;\\
            Calculate Gradients $\frac{\partial\mathcal{L}}{\partial\Theta}$;\\
            Optimize $\Theta$ via the Adam Optimizer
        }
    }
    \BlankLine
    \textbf{Context Aware Pixel Generation:}\\
    $\CAPixel= \PCFunc(Y,\NBPixel)$\\
    \Return $\CAPixel$
    
    \caption{Training of the \acrfull{an}}\label{alg:PC}
\end{algorithm}

\begin{algorithm}

    \SetKwFunction{RP}{RP}
    \SetKwProg{Fn}{Function}{:}{}
    \SetKwInOut{Input}{Input}
    \SetKwInOut{Output}{Output}
    \Input{\\
    
    Contextualized Pixels: $\CAPixel = \{\Bar{y_i}\}_{i=1}^N$\\
    Endmember Ensembles:$\{\Eensemble_i\}_{i=1}^M$\\
    Observed Pixels: $Y = \{y_i\}_{i=1}^N$,\\
    Learning rate: $\alpha$, Batch Size : $D$,\\
    $\text{Epochs}_\text{Stage 1}$, $\text{Epochs}_\text{Stage 2}$,\\
    Hyperparameters : \\
    \Indp
    $\lambda_{1}, \lambda_{2}, \lambda_{3}, \lambda_{4},$\\
    
    \Indm
    }
    \Output{\\
        Predicted Abundances: $\widehat{A}$,\\
        Predicted Signatures: $\widehat{S}$\\
    }
    
    \BlankLine
    Initialize \acrshort{ap} Model Parameters: $\Psi$;\\
    Initialize Endmember Query Vectors: $\Omega$;\\
    Initialize \acrshort{sp} Multihead-Attention Parameters: $\Phi$;\\
    \BlankLine
    
    \textbf{Training Stage 1:}\\

    \For{$\textnormal{epoch}=1$ \KwTo $\textnormal{Epochs}_\textnormal{Stage 1}$}{
        \For{$\textnormal{batch}=1$ \KwTo $\left\lceil \frac{N}{D}\right\rceil$}{
            Sample $D$ pixels ($\CAPixel_{\textnormal{batch}}$),($Y_{\textnormal{batch}}$) from $(\CAPixel)$,$Y$;\\
            Predict Abundances:\\
                \Indp
                    $\widehat{A}_{\text{batch,epoch}} = \APFunc(\CAPixel_{\textnormal{batch}})$;\\
                \Indm
            Predict Signatures:\\
                \Indp
                    $\widehat{S}_{\text{batch,epoch}} = \SPFunc(\{\Eensemble_i\}_{i=1}^M)$;\\
                \Indm
            Predicted Pixels:\\
                \Indp
                    $\widehat{Y}_{\text{batch,epoch}} = \widehat{S}_{\text{batch,epoch}}\;\widehat{A}_{\text{batch,epoch}}$;\\
                \Indm
            Loss:\\
            $\mathcal{L}_1 = [\lambda_{1}\mathcal{L}_{MSE}(\widehat{Y}_{\text{batch,epoch}},Y_{\textnormal{batch}})+
            \lambda_{2}\mathcal{L}_{SAD}(\widehat{Y}_{\text{batch,epoch}},Y_{\textnormal{batch}})+
            \lambda_{3}\mathcal{L}_{Non-Neg}(\widehat{S}_{\text{batch,epoch}})]$\\
            \BlankLine 
            Calculate Gradients $\frac{\partial\mathcal{L}_1}{\partial\Psi},\frac{\partial\mathcal{L}_1}{\partial\Omega}$;\\
            Optimize $\Psi,\Omega$ via the Adam Optimizer
            
        }
    }
    \BlankLine
    
    \textbf{Calculating the Control Volume:}\\
    $\widehat{S} = \text{SP}_{\Phi_a,\Phi_b}(\{\Eensemble_i\}_{i=1}^M)$\\
    Control Volume: $V_T$ = $\mathcal{V}(\widehat{S}$);\\
    \BlankLine
    
    \textbf{Training Stage 2:}\\
    
    \For{$\textnormal{epoch}=1$ \KwTo $\textnormal{Epochs}_\textnormal{Stage 2}$}{
        \For{$\textnormal{batch}=1$ \KwTo $\left\lceil \frac{N}{D}\right\rceil$}{
            $\dots$\\
            Same as Training Stage 1 Line 7 to 15\\
            $\dots$\\
            $\mathcal{L} = \mathcal{L}_1+\lambda_4\mathcal{L}_{Min-Vol}(\widehat{S},V_T)$\\
            Calculate Gradients $\frac{\partial\mathcal{L}}{\partial\Psi},\frac{\partial\mathcal{L}}{\partial\Phi}$ $\frac{\partial\mathcal{L}}{\partial\Omega}$;\\
            Optimize $\Psi,\Phi,\Omega$ via the Adam Optimizer
        }
    }
    \BlankLine
    \textbf{Abundance and Endmember Signature Prediction:}\\
    $\widehat{A} = \APFunc(\CAPixel)$;\\
    $\widehat{S} = \SPFunc(\{\Eensemble_i\}_{i=1}^M)$;\\
    \BlankLine
    \Return $\widehat{A},\widehat{S}$
    
    \caption{Training the \acrfull{ap} and \acrfull{sp}}\label{alg:AP_SP}
\end{algorithm}
\subsection{Training Process}
\label{training process}


To provide an overview of the training process, the \acrfull{an} is initially trained to generate Contextualized Pixels ($\CAPixel$), as demonstrated in Algorithm \ref{alg:PC}. These Contextualized Pixels are subsequently utilized in the next phase, where the \acrfull{ap} and \acrfull{sp} are trained together. After training the \acrshort{ap} and \acrshort{sp}, the final estimated abundance and endmember signatures are predicted, as illustrated in Algorithm \ref{alg:AP_SP}. The detailed procedures of these steps will be discussed in the following sections.

\subsubsection{Training the \acrfull{an}}


As illustrated in Algorithm \ref{alg:PC}, the \acrshort{an} is trained to minimize the Mean Square Error (MSE). As discussed in Section \ref{sec:PC}, this training approach ensures that the output is similar to the original pixel (Observed pixel). However, because the output is generated by combining the transformed neighborhood pixels (which are passed as the Value parameter), and the weights are conditioned on the relationship between the neighborhood pixels and the original pixel, the context of the neighborhood is effectively incorporated into the output.

This means that the \acrshort{an} integrates relevant information from the surrounding pixels, enhancing the accuracy and contextual relevance of the output. This neighborhood context injection is crucial for capturing the local spatial relationships and improving the overall performance of the model.


\subsubsection{Training of the \acrfull{ap} and \acrfull{sp}}
The algorithm for the training of \acrshort{ap} and \acrshort{sp} is given in Algorithm \ref{alg:AP_SP}. As previously mentioned, the model's convergence is achieved by minimizing the reconstruction loss of the \acrshort{hs} image reconstructed from the predicted endmember abundances and signatures. However, since the \acrshort{ap} is initially untrained, its predictions may be incorrect, potentially leading the \acrshort{sp} astray and causing the model to converge to a local minimum. On the other hand, the \acrshort{sp}, supported by a candidate ensemble, has the potential to produce accurate endmembers close to the true values with the right parameters. Therefore, it is advantageous to initially guide the abundance predictor in the correct direction for proper convergence.

To achieve this, the training process is divided into two stages. In the first stage (See Algorithm \ref{alg:AP_SP} line 4), certain restrictions are placed on the parameters of the \acrshort{sp}, enabling it to predict endmembers that are reasonably accurate using the ensemble information provided. This helps guide the \acrshort{sp} in the correct direction. In the second stage (See Algorithm \ref{alg:AP_SP} line 23), these restrictions are removed, allowing the model parameters to train freely and converge to the correct values. The following paragraphs will explain these stages in more detail.

In the first training stage, the parameters for the \acrshort{ap} are randomly initialized. For the \acrshort{sp}, the linear layers ($\Phi$) are initially set to identity matrices and kept constant during this stage, as shown in Algorithm \ref{alg:AP_SP} lines 16 and 17. This ensures that the ensemble is not modified initially and allows the model to simply weight the ensemble components when combining them. Although this is suboptimal, it provides initial guidance to the \acrshort{ap}. The Query parameters ($\Omega$), on the other hand, are randomly initialized and allowed to train, enabling the model to intelligently combine the ensemble. This stage is conducted for a specified number of epochs.

The algorithm then proceeds to the second stage of the training process. In this stage, the linear layers ($\Phi$) are allowed to train freely, as shown in pseudocode \ref{alg:AP_SP} lines 30 and 31. At this point, the \acrshort{sp} has full flexibility to train the ensemble. However, adding the Min-Vol constraint on the predicted endmembers is crucial in this stage to prevent the endmember simplex volume from increasing far beyond the data simplex. Therefore, an additional loss function is incorporated into the previous reconstruction loss (refer line 29 of Algorithm \ref{alg:AP_SP}). This loss penalizes cases where the volume of the predicted endmember simplex exceeds a certain threshold. The details of the volume calculation are explained in Section \ref{sec:min_vol_calc}. The threshold volume is the volume of the predicted endmember simplex at the end of Stage 1 training (refer to line 20 of Algorithm \ref{alg:AP_SP}).

Finally, the trained models are used to predict the endmember abundances and signatures as can be seen in line 34 of Algorithm \ref{alg:AP_SP}.

\section{EXPERIMENTS}
\label{experiments}

\begin{table*}[!t]
\centering
\caption{Optimum Hyperparameters of Proposed Method for Each Dataset}
\begin{tabular}{c|c|c|c|c|c|c|c}
\toprule
Dataset &Epochs (\acrshort{an}) & $\text{Epochs}_\text{Stage 1}$ & $\text{Epochs}_\text{Stage 2}$ & SAD & MSE & Min-Vol & Non-neg\\
\toprule
Samson	&100 &1000	&500	&1.125	&1 &0.0025 &1e-8\\
\hline
Jasper	&200	&1000	&350	&1e-5 &1 &5e-5 &1e-6\\
\hline
Urban	&100	&1000	&0	&0 &1 &0 &1e-8\\
\hline
Synthetic	&200	&1000	&0	&1.125 &1 &0 &1e-8\\
\bottomrule
\end{tabular}
\label{table: hyperparameter}
\end{table*}
\begin{figure*}[!t]
    \centering
    \includegraphics[width=0.85\textwidth]{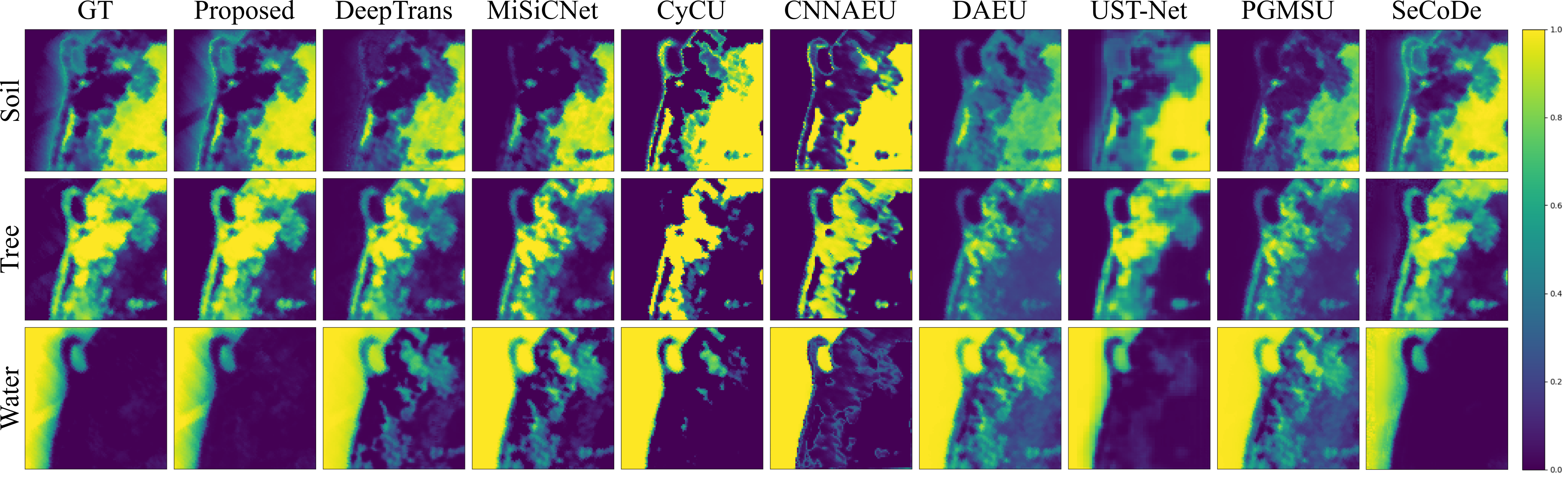}
    \caption{Abundance Maps Obtained by the Different Unmixing Algorithms for Samson Dataset}
    \label{samson_sign}
\end{figure*}
\begin{figure*}[!t]
    \centering
    \includegraphics[width=0.85\textwidth]{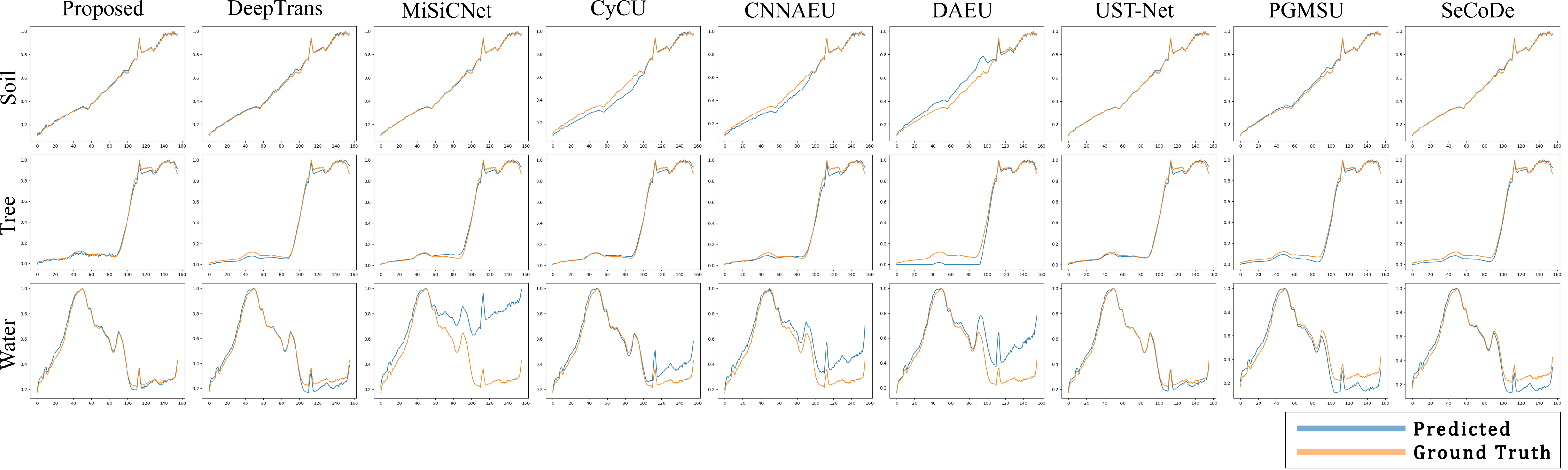}
    \caption{Endmember Spectral Signatures Obtained by the Different Unmixing Algorithms for Samson Dataset}
    \label{samson_abd}
\end{figure*}

\begin{figure}[!t]
\label{rgb_dataset}
\includegraphics[width=\linewidth]{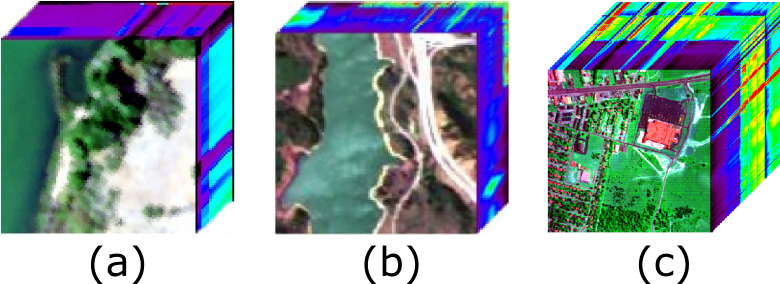}
  \caption{RGB Images of the Real Datasets: (a) Samson, (b) Jasper-Ridge and, (c) Urban }
\end{figure}

For the performance evaluation, three real datasets \cite{rslab} and one synthetic dataset have been used. Real datasets are Samson, Jasper-Ridge, and Urban. The RGB images of the real datasets are shown in Figure \ref{rgb_dataset}. For the results comparison the State-Of-The-Art algorithms such as DeepTrans \cite{transformer}, MiSiCNet \cite{misicnet}, CyCU \cite{cycu}, CNNAEU \cite{cnnaeu}, DAEU \cite{ae}, USTNet \cite{UST}, PGMSU \cite{PGMSU}, and SeCoDE \cite{Secode} has been used.


\subsection{Datasets}

\subsubsection{Samson}
Samson dataset is a widely used dataset in HS unmixing. It consists of three endmembers namely, tree, sand, and water. The number of bands present in the dataset is 156 covering from 401nm to 889nm with a resolution up to 3.13nm.  The original image has 952x952 pixels. For simplicity, a patch of 95x95 has been used. 

\subsubsection{Jasper Ridge}
The Jasper Ridge dataset is another commonly used dataset. The original image has 512x614 pixels. As it is too complex to get the ground truth, a subimage of 100x100 is considered. In the original image, there are 224 channels from 380nm to 2500nm with a resolution of up to 9.46nm. After removing 1-3, 108-112, 154-166, and 220-224 bands due to dense water vapor and atmospheric effects there are 198 channels in the HS image. The endmembers in this HS image are road, soil, water, and tree.

\subsubsection{Urban}

The Urban dataset is extensively utilized in the \acrshort{hs}  unmixing field. The original images contain 210 bands, but bands 1-4, 76, 87, 101-111, 136-153, and 198-210 are excluded due to dense water vapor and atmospheric effects. As a result, the final image consists of 162 channels with dimensions of 307x307 pixels. The channels cover a wavelength range from 400nm to 2500nm, with a bandwidth resolution of 10nm. This dataset includes three ground truths, each containing 4, 5, and 6 endmembers. These endmembers are asphalt, grass, tree, roof, dirt, and metal.

\subsubsection{Synthetic}
This dataset consists of images with dimensions of 90x90 pixels, featuring four endmembers and 144 spectral bands. The endmember signatures were selected from the existing synthetic dataset as referenced in \cite{weebly}. The generation of abundance maps was inspired by "Worley Noise" \cite{Worley_Noise} and the use of Dirichlet distribution \cite{diri} which ensures that both the \acrshort{anc} and the \acrshort{asc} are satisfied in the abundance maps.

\begin{table*}[!t]
\label{samson_compare}
\centering
\caption{Unmixing Performance Comparison for Different \acrfull{hs} Unmixing Algorithms for Samson Dataset}
\footnotesize

\begin{tabular}{c |c| c| c| c| c| c| c| c| c| c}
\Xhline{2pt}
Error & Endmember & Proposed & DeepTrans & MiSiCNet & CyCU & CNNAEU & DAEU & USTNet & PGMSU & SeCoDe\\
\Xhline{2pt}
\multirow{4}{4em}{RMSE} & Soil 
&\textbf{0.0442\textsuperscript{1}} &0.1717 &0.1818 &0.2372 &0.1924 &0.3042 &0.1068\textsuperscript{3} &0.2191 &0.0744\textsuperscript{2} \\
&Tree 
&\textbf{0.0266\textsuperscript{1}} &0.1720 &0.1791 &0.2635 &0.1846 &0.5845 &0.1086\textsuperscript{3} &0.2442 &0.0705\textsuperscript{2} \\
& Water 
&\textbf{0.0258\textsuperscript{1}} &0.1715 &0.3131 &0.1918 &0.1744 &0.2462 &0.0922\textsuperscript{3} &0.3577 &0.0284\textsuperscript{2} \\
&Average	
&\textbf{0.0333\textsuperscript{1}} &0.1717 &0.2332 &0.2327 &0.1839 &0.4061 &0.1028\textsuperscript{3} &0.2802 &0.0614\textsuperscript{2} \\
[1ex] 
\hline

\multirow{4}{4em}{SAD} & Soil 
&0.0117 &0.0108	&0.0103\textsuperscript{3} &0.0649 &0.0512 &0.9611 &\textbf{0.0081\textsuperscript{1}} &0.0207 &0.0102\textsuperscript{2} \\
& Tree
&0.0309\textsuperscript{2}	&0.0425	&0.0354	&\textbf{0.0275\textsuperscript{1}} &0.0408 &0.7004 &0.0311\textsuperscript{3} &0.0495 &0.0490 \\
& Water 
&\textbf{0.0326\textsuperscript{1}} &0.0568\textsuperscript{3} &0.4011 &0.1071 &0.1544 &1.4270 &0.0381\textsuperscript{2}  &0.1299 &0.1050 \\
& Average 
&\textbf{0.0250\textsuperscript{1}} &0.0367\textsuperscript{3} &0.1489 &0.0665 &0.0821 &1.0295 &0.0258\textsuperscript{2} &0.0667 &0.0547 \\
[1ex] 
\Xhline{2pt}
\end{tabular}

\label{table: comparision_of_algorithms_samson}
\end{table*}
\subsection{Experiment Setup}

\subsubsection{Evaluation Metrics}

For the performance comparison, the \acrfull{rmse} and \acrfull{sad} values have been compared for each endmember as well as for the average value. \acrshort{rmse} reflects the abundance maps accuracy. It calculates the root mean squared error between the generated abundance maps and the ground truth abundance maps. This is depicted in equation \ref{RMSE}. \acrshort{sad} evaluates the accuracy of the endmember signatures. The comparing parameter in \acrshort{sad} is the angle distance between the ground truth signature and the generated signature. The is shown in equation \ref{SAD}.

\begin{equation}
RMSE(\widehat{\mathbf{a}},\mathbf{a}) = \sqrt{\frac{1}{N}\sum\limits_{i=1}^N (\widehat{\mathbf{a_i}} - \mathbf{a_i})^2} 
\label{RMSE}
\end{equation}

\begin{align}
    SAD(\mathbf{\widehat{x}},\mathbf{x}) = \arccos \left({\frac{\mathbf{\widehat{x}^T} \mathbf{x}}{||\mathbf{\widehat{x}}||\;||\mathbf{x}||}}\right)
\label{SAD}   
\end{align}

\subsubsection{Endmember Extraction Algorithm for Fusion}

As discussed in the methodology section, \acrshort{sp} uses an ensemble of endmember signatures extracted by different EEAs used and predict the endmember spectral signatures. Technically, for creating the ensemble any number of EEAs could be used. In this paper, 3 key EEAs are used for demonstration. These are VCA \cite{vca}, N-FINDR \cite{nfindr}, and ATGP.

Initially, every combination of these 3 EEAs has been used to find the best combination that gives the best \acrshort{rmse} and \acrshort{sad} results. For the Samson dataset, the obtained results are tabulated in Table \ref{table: EEA combinations}.

It can be clearly seen that fusing all EEAs gives the best \acrshort{rmse} and \acrshort{sad}. Therefore that combination is used for following sections.

\subsubsection{Hyperparameter Settings}

The used hyperparameters in the proposed \acrshort{fs} algorithm are listed below: The layers for each section in taken to align with the conceptual idea that has been discussed in the methodology. Adam optimizer is used for each training stage and the learning rate is taken as $10^{-4}$. The input batch size is 400. 
Apart from these, the optimum weights of the SAD, MSE, Non-negative and Min-vol losses in the loss function were different from dataset to dataset. Also, the number of epochs for each training stage differs according to the dataset. The optimum values are given in the Table \ref{table: hyperparameter} for each dataset. 



\begin{figure*}[!t]
\includegraphics[width=\linewidth]{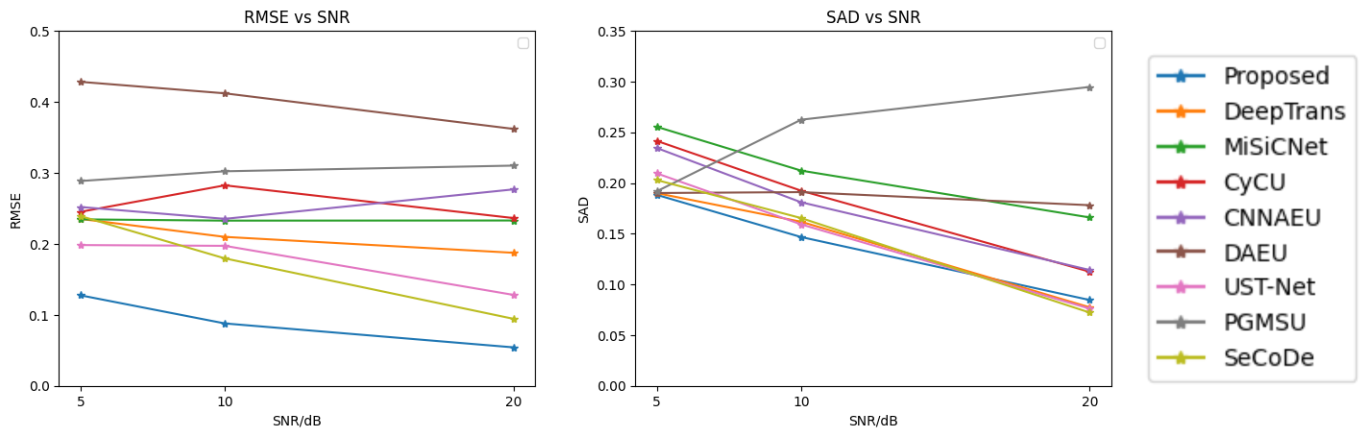}
\caption{Results of the Different Unmixing Algorithms for Samson Dataset Under 20dB, 10dB and 5dB SNRs}
\label{noise_performance_of_each_algo}
\end{figure*}
\subsection{Experiment with Samson Dataset}

\begin{table*}[!t]
\centering
\caption{Performance of Proposed Method Under Different Levels of Noise for Samson Dataset}
\footnotesize

\begin{tabular}{c |c| c| c| c| c| c| c| c| c}
        \Xhline{2pt}
        \multicolumn{2}{c|}{}& \multicolumn{2}{c|}{Original Image} & \multicolumn{2}{c|}{20dB SNR} & \multicolumn{2}{c|}{10dB SNR} & \multicolumn{2}{c}{5dB SNR}\\
        \hline
        Error & Endmember & Mean & Std & Mean & Std & Mean & Std & Mean & Std\\
        \Xhline{2pt}
        \multirow{4}{4em}{RMSE} & Soil & 0.0632 & 0.0093 & 0.0714 & 0.0103 & 0.8790 & 0.0100 & 0.1181 & 0.0057\\
        & Tree & 0.0361 & 0.0058 & 0.0575 & 0.0089 & 0.0854 & 0.0076 & 0.1282 & 0.0039\\
        & Water & 0.0351 & 0.0064 & 0.0497 & 0.0059 & 0.1034 & 0.0077 & 0.1397 & 0.0049\\
        &Average	&0.0467	&0.0069 &	0.0604&	0.0064&	0.0927&	0.0072 &	0.1290	&0.0036 \\
        [1ex] 
        \hline 
        \multirow{4}{4em}{SAD} & Soil 
&0.0123	&0.0017	&0.0332	&0.0050	&0.0830	&0.0057	&0.1218	&0.0068\\
& Tree
&0.0328	&0.0021	&0.0526	&0.0047	&0.0982	&0.0190	&0.1570	&0.0112\\
& Water 
&0.0330	&0.0019	&0.1653	&0.0070	&0.2721	&0.0144	&0.2861	&0.0051\\
& Average 
&0.0260	&0.0009	&0.0837	&0.0030	&0.1511	&0.0055	&0.1883	&0.0028\\
[1ex] 
        \Xhline{2pt}
\end{tabular}
\label{table: noise with different algos}
\end{table*}

\begin{table*}[!t]
\centering
\caption{Unmixing Performance Comparison for Different EEA Combinations for Samson Dataset.}
\footnotesize

\begin{tabular}{c |c| c| c| c| c| c| c| c}
\Xhline{2pt}
Error & Endmember & VCA  & NFINDR & ATGP & VCA + ATGP & VCA + NFINDR & ATGP + NFINDR & All EEAs\\
\Xhline{2pt}
\multirow{4}{4em}{RMSE} & Soil 
&0.0645	&0.1030	&0.1605	&0.0552	&0.0888	&0.0712	&\textbf{0.0442}\\
&Tree 
&0.0334	&0.0908	&0.1192	&0.0317	&0.0701	&0.0446	&\textbf{0.0266}\\
& Water 
&0.0406	&0.0374	&0.1199	&0.0317	&0.0443	&0.0414	&\textbf{0.0258}\\
&Average	
&0.0481	&0.0822	&0.1346	&0.0410	&0.0701	&0.0541	&\textbf{0.0333}\\
[1ex] 
\hline

\multirow{4}{4em}{SAD} & Soil 
&0.0203	&0.0764	&0.1568	&0.0132	&0.0560	&0.0300	&\textbf{0.0117}\\
& Tree
&0.0328	&0.0334	&0.0387	&0.0325	&0.0335	&0.0352	&\textbf{0.0309}\\
& Water 
&0.0391	&0.0359	&0.1440	&0.0311	&0.0416	&0.0326	&\textbf{0.0326}\\
& Average 
&0.0307	&0.0485	&0.1132	&0.0256	&0.0437	&0.0326	&\textbf{0.0250}\\
[1ex] 
\Xhline{2pt}
\end{tabular}

\label{table: EEA combinations}
\end{table*}
\subsubsection{Perfomance Comparison with the State-Of-The-Art Algorithms}

For the Samson dataset, the abundance maps and the endmember signatures are shown in Figure \ref{samson_sign} and Figure \ref{samson_abd} respectively. Numerical value comparison is shown in Table \ref{table: comparision_of_algorithms_samson} and the time taken for training and the prediction are tabulated in \ref{table: time_cost}.

In terms of \acrshort{rmse}, the proposed algorithm demonstrates superior performance, achieving the best results across all endmembers. Except for the \acrshort{sad} values associated with soil and tree endmembers, where USTNet and CyCU algorithms exhibit slightly superior performance, the proposed method demonstrates improved results for the water endmember and, on average, outperforms the other algorithms. 

\subsubsection{Effect of Noise}
In this section, the effect of various levels of noise on each algorithm is analyzed. To perform this, the original Samson \acrshort{hs} image with added noise is used. The noise is added to generate images with Signal-to-Noise-Ratios (SNRs) of 20 dB, 10 dB, and 5 dB. The results obtained from each algorithm for these images are shown in Figure \ref{noise_performance_of_each_algo}.

The best results for \acrshort{rmse} were by the proposed algorithm for each noise level with a noticeable margin. Considering \acrshort{sad} values, the proposed method gives significantly better performance for 10dB noise level and comparable results for other noise levels.

Since the addition of noise is an inherently random process, variations in results can be expected under different noise conditions. To analyze this variability, we conducted a detailed performance evaluation of the proposed algorithm by running it 10 times for each noise level. The mean and standard deviation values for each error metric are presented in Table \ref{table: noise with different algos}. This table illustrates the reproducibility of the results under varying noise conditions.

\subsection{Experiment with Jasper-Ridge Dataset}

\subsubsection{Perfomance Comparison with the State-Of-The-Art Algorithms}

\begin{figure*}[!t]
    \centering
    \includegraphics[width=0.75\textwidth]{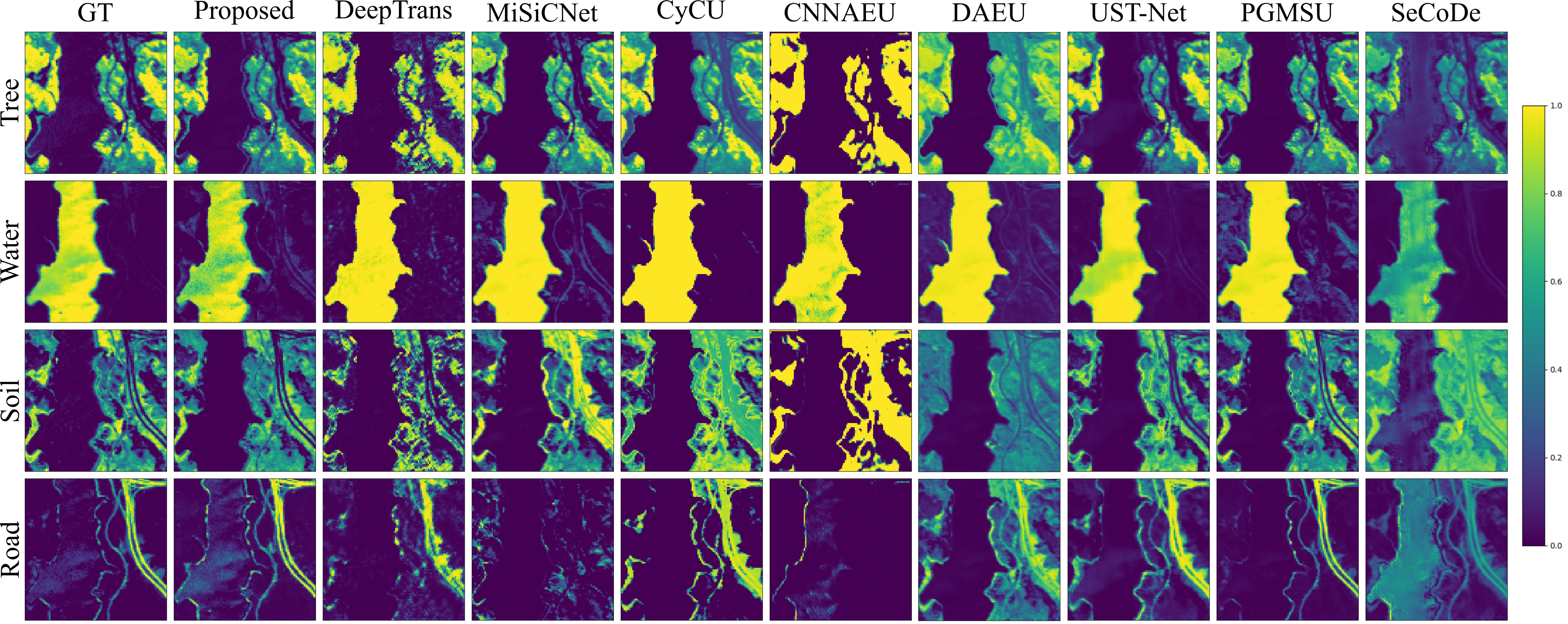}
    \caption{Abundance Maps Obtained by the Different Unmixing Algorithms for Jasper-Ridge Dataset}
    \label{jasper_sign}
\end{figure*}

\begin{figure*}[!t]
    \centering
    \includegraphics[width=0.85\textwidth]{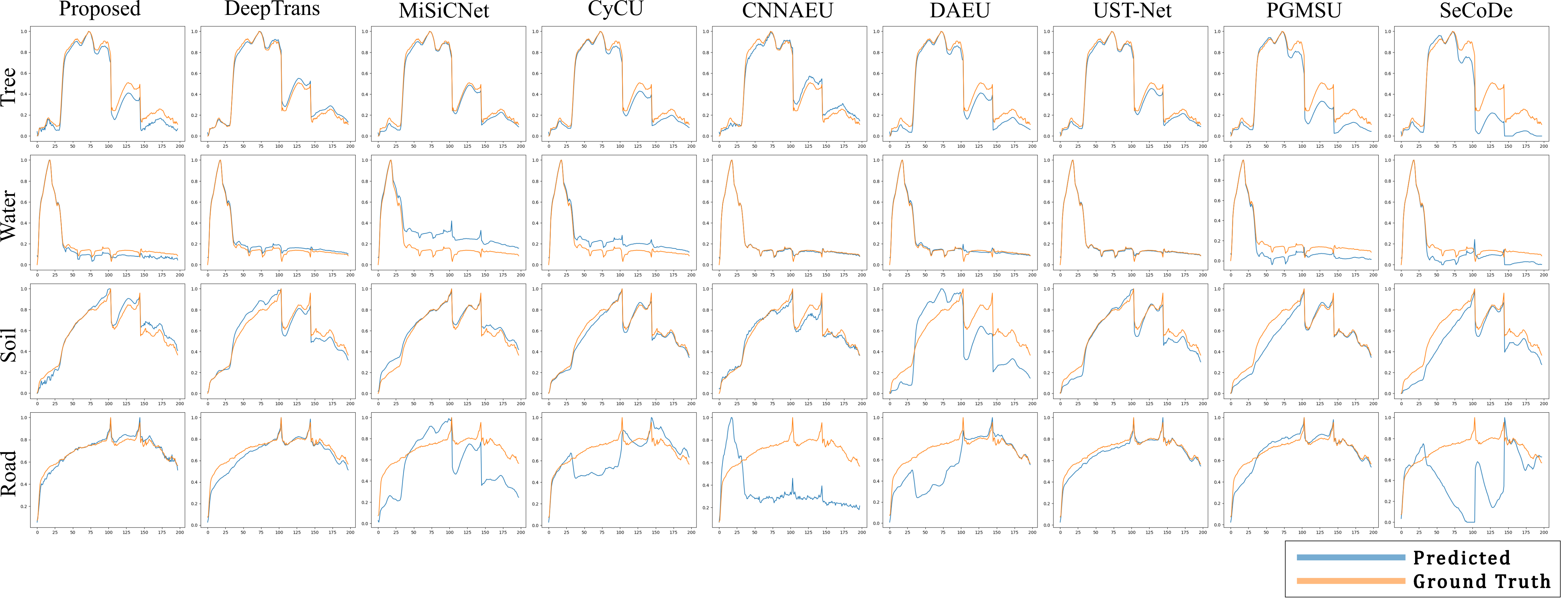}
    \caption{Endmember Spectral Signatures Obtained by the Different Unmixing Algorithms for Jasper-Ridge Dataset}
    \label{jasper_abd}
\end{figure*}

For the Jasper-Ridge dataset, the ground truth abundance maps and the predicted abundance maps for each algorithm are shown in Figure \ref{jasper_abd}. The comparison of the predicted signatures with the ground truth for each algorithm is given in Figure \ref{jasper_sign}. A comparison of the different error metrics for each algorithm is given in the table \ref{table: comparision_of_algorithms_jasper} and the times taken for training and the prediction are tabulated in \ref{table: time_cost}.


When considering the \acrshort{rmse} of the individual endmember abundance, the proposed algorithm shows the best performance for tree and soil abundance values while showing the second-best performance for the other endmembers. Furthermore, the proposed algorithms have the best average \acrshort{rmse} value. When considering \acrshort{sad} values, the proposed algorithm shows the best result for soil and road endmembers. 

\begin{table*}[!t]
\centering
\caption{Time Taken by Each Dataset for Training and Prediction}
\begin{tabular}{c|c|c|c|c|c}
\Xhline{2pt}
Dataset	&\acrshort{an} Training (s)	&\acrshort{ap}+\acrshort{sp}: Stage 1 (s) &\acrshort{ap}+\acrshort{sp}: Stage 2 (s) &AP Prediction (s)	&SP Prediction (s)\\
\Xhline{2pt}
Samson	&117.07	&440.53	&245.71	&0.0285	&0.0031\\
\hline
Jasper-Ridge &254.04	&662.01	&243.18	&0.0495	&0.0035\\
\hline
Urban &939.23	&5916.12	&0	&0.6328	&0.0056\\
\hline
Synthetic &168.21	&453.88	&0	&0.0427	&0.0036\\
\Xhline{2pt}
\end{tabular}
\label{table: time_cost}
\end{table*}

\subsubsection{Performance of the Signature Predictor}
\label{performance of SP}


\begin{table*}[!t]
\centering
\caption{Unmixing Performance Comparison for Different \acrfull{hs} Unmixing Algorithms for Jasper Dataset.}
\footnotesize

\begin{tabular}{c |c| c| c| c| c| c| c| c| c| c}
\Xhline{2pt}
Error & Endmember & Proposed & DeepTrans & MiSiCNet & CyCU & CNNAEU & DAEU &USTNet &PGMSU &SeCoDe\\
\hline 
\multirow{4}{4em}{RMSE} & Tree 
&\textbf{0.0822\textsuperscript{1}} &0.1388 &0.0954\textsuperscript{2} &0.1082 &0.2172 &0.1633 &0.1036\textsuperscript{3} &0.1087 &0.2406\\
& Water 
&0.0807\textsuperscript{2} &0.0901\textsuperscript{3} &0.1141 &0.1044 &0.1454 &0.1272 &\textbf{0.0375}\textsuperscript{1} &0.1211 &0.2000 \\
& Soil 
&\textbf{0.0776\textsuperscript{1}} &0.2068 &0.2216 &0.1690 &0.3256 &0.2099 &0.1585\textsuperscript{3}  &0.0808\textsuperscript{2} &0.2727\\
& Road 
&0.1048\textsuperscript{2} &0.1545 &0.2487 &0.1362\textsuperscript{3} &0.2383 &0.2165 &0.1493 &\textbf{0.0582\textsuperscript{1}}  &0.1773\\
&Average	
&\textbf{0.0854\textsuperscript{1}} &0.1533 &0.1824 &0.1320 &0.2404 &0.1829 &0.1220\textsuperscript{3}	&0.0954\textsuperscript{2}	&0.2257\\
[1ex] 
\hline

\multirow{4}{4em}{SAD} & Tree 
&0.1222	&0.0531\textsuperscript{3}	&\textbf{0.0434\textsuperscript{1}} &0.0662 &0.0818 &0.1807 &0.0479\textsuperscript{2}	&0.1657	&0.2861\\
& Water 
&0.1355 &0.0743 &0.2894 &0.1739 &0.0334\textsuperscript{2} &0.0639\textsuperscript{3} &\textbf{0.0322\textsuperscript{1}}	&0.2582	&0.2470\\
& Soil 
&\textbf{0.0427\textsuperscript{1}} &0.0915 &0.0662 &0.0477\textsuperscript{2} &0.0565\textsuperscript{3} &0.3453 &0.0770	&0.1166	&0.1327\\
& Road 
&\textbf{0.0421\textsuperscript{1}} &0.0698\textsuperscript{3}	&0.3295	&0.2071	&0.5969	&0.4180 &0.0457\textsuperscript{2}	&0.0901	&0.5797\\
&Average	
&0.0870\textsuperscript{3} &0.0722\textsuperscript{2} &0.1821	&0.1237	&0.1921	&0.2520 &\textbf{0.0507\textsuperscript{1}}	&0.1577	&0.3114\\
[1ex] 
\Xhline{2pt}
\end{tabular}

\label{table: comparision_of_algorithms_jasper}
\end{table*}

\begin{figure*}[!t]
    \centering
    \includegraphics[width=0.75\textwidth]{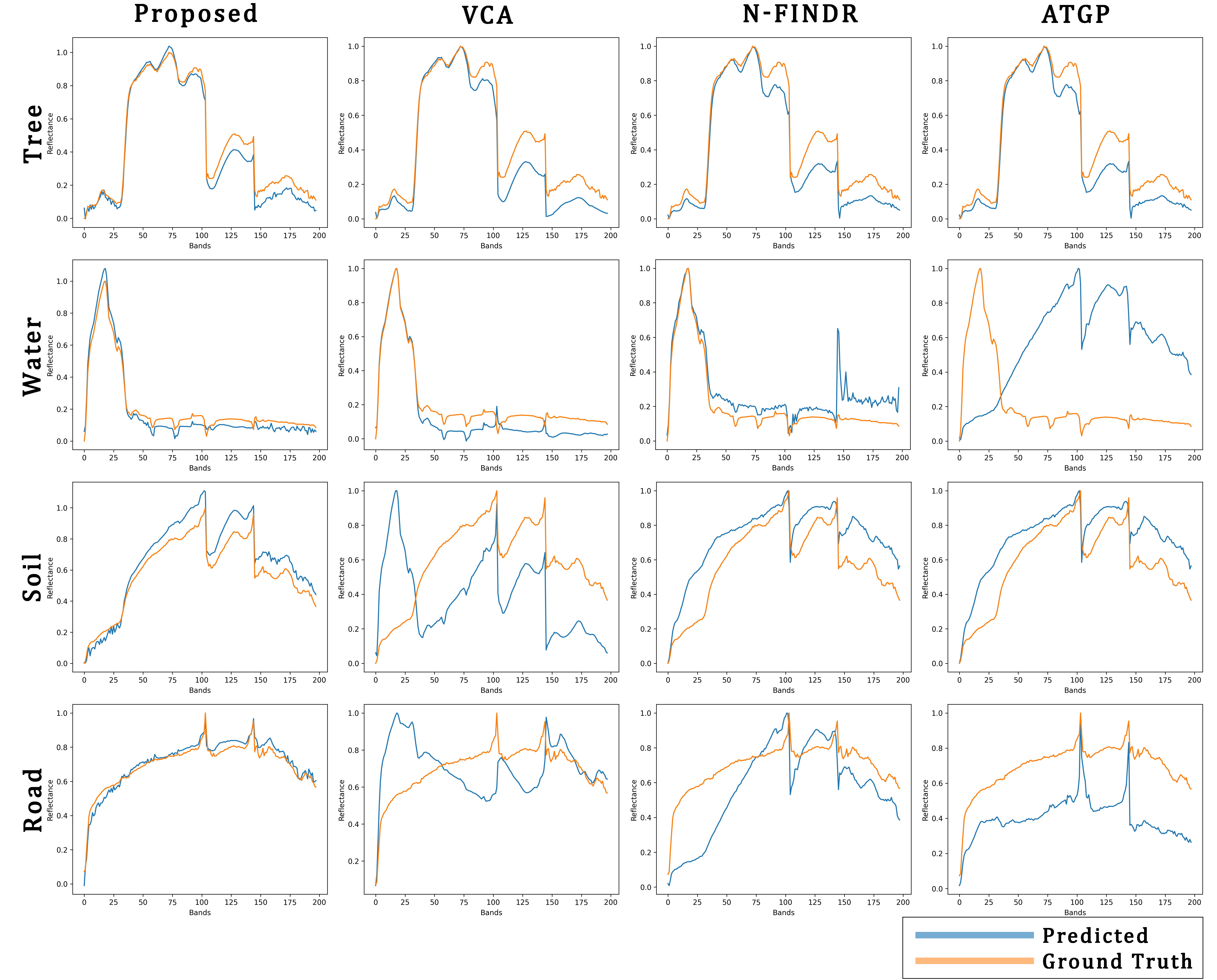}
    \caption{Endmember Signatures Obtained by \acrshort{fs} and Different \acrshort{eea}s for Jasper-Ridge Dataset}
    \label{eea_comparison}
\end{figure*}

\begin{table*}[!t]
\centering
\caption{Unmixing Performance Comparison for Different \acrfull{hs} Unmixing Algorithms for Urban Dataset.}
\footnotesize

\begin{tabular}{c |c| c| c| c| c| c| c| c| c| c}
\Xhline{2pt}
Error & Endmember & Proposed & DeepTrans & MiSiCNet & CyCU & CNNAEU & DAEU & USTNet & PGMSU & SeCoDe\\
\Xhline{2pt}
\multirow{4}{4em}{RMSE} & Asphalt 
&0.0507\textsuperscript{2}	&0.1242	&\textbf{0.0423\textsuperscript{1}}	&0.2803	&0.2461	&0.3530	&0.1442	&0.0825\textsuperscript{3}	&0.2023 \\
&Grass 
&\textbf{0.1174\textsuperscript{1}}	&0.2911	&0.1677	&0.3932	&0.3342	&0.3731	&0.2403	&0.1658\textsuperscript{3}	&0.1382\textsuperscript{2} \\
& Tree 
&0.1005\textsuperscript{3}	&0.0899\textsuperscript{2}	&0.1334	&0.2403	&0.2620	&0.3463	&0.1115	&0.1361	&\textbf{0.0769\textsuperscript{1}} \\
& Roof 
&0.0408\textsuperscript{2}	&0.0773	&\textbf{0.0243\textsuperscript{1}}	&0.1369	&0.2124	&0.1707	&0.1252	&0.0662	&0.0630\textsuperscript{3} \\
& Metal 
&\textbf{0.0495\textsuperscript{1}}	&0.3227	&0.0566\textsuperscript{2}	&0.2484	&0.1321	&0.1166	&0.1525	&0.1063	&0.0622\textsuperscript{3} \\
& Dirt 
&\textbf{0.0338\textsuperscript{1}}	&0.1004	&0.0365\textsuperscript{2}	&0.1726	&0.2055	&0.2062	&0.1190	&0.0607\textsuperscript{3}	&0.2823 \\
&Average	
&\textbf{0.0727\textsuperscript{1}}	&0.1951	&0.0938\textsuperscript{2}	&0.2586	&0.2400	&0.2796	&0.1550	&0.1097\textsuperscript{3}	&0.1599 \\
[1ex] 
\hline

\multirow{4}{4em}{SAD} & Asphalt 
&\textbf{0.0000\textsuperscript{1}}	&0.0203	&0.0075\textsuperscript{3}	&0.5444	&0.0425	&0.1053	&0.0117	&\textbf{0.0000\textsuperscript{1}}	&0.2551\\
& Grass
&\textbf{0.0000\textsuperscript{1}}	&0.0068\textsuperscript{3}	&0.0952	&0.4203	&0.0886	&0.4261	&0.0114	&\textbf{0.0000\textsuperscript{1}}	&0.2038\\
& Tree 
&\textbf{0.0000\textsuperscript{1}}	&0.0045\textsuperscript{3}	&0.0051	&0.1200	&0.0139	&0.1247	&0.0053	&\textbf{0.0000\textsuperscript{1}}	&0.2115\\
& Roof 
&\textbf{0.0000\textsuperscript{1}}	&0.0358	&0.0331	&0.1032	&0.0697	&0.0969	&0.0308\textsuperscript{3}	&\textbf{0.0000\textsuperscript{1}}	&0.2957\\
& Metal 
&\textbf{0.0000\textsuperscript{1}}	&0.3725	&0.1948	&0.1927\textsuperscript{3}	&0.5831	&0.3278	&0.6373	&\textbf{0.0000\textsuperscript{1}}	&0.2630\\
& Dirt 
&\textbf{0.0000\textsuperscript{1}}	&0.0550	&0.0178	&0.0714	&0.0289	&0.3020	&0.1220	&\textbf{0.0000\textsuperscript{1}}	&0.0099\textsuperscript{3}\\
& Average 
&\textbf{0.0000\textsuperscript{1}}	&0.0825	&0.0589\textsuperscript{3}	&0.2420	&0.1378	&0.2305	&0.1364	&\textbf{0.0000\textsuperscript{1}}	&0.2065\\
[1ex] 
\Xhline{2pt}
\end{tabular}

\label{table: comparision_of_algorithms_urban}
\end{table*}
\begin{figure*}[!t]
    \centering
    \includegraphics[width=0.75\textwidth]{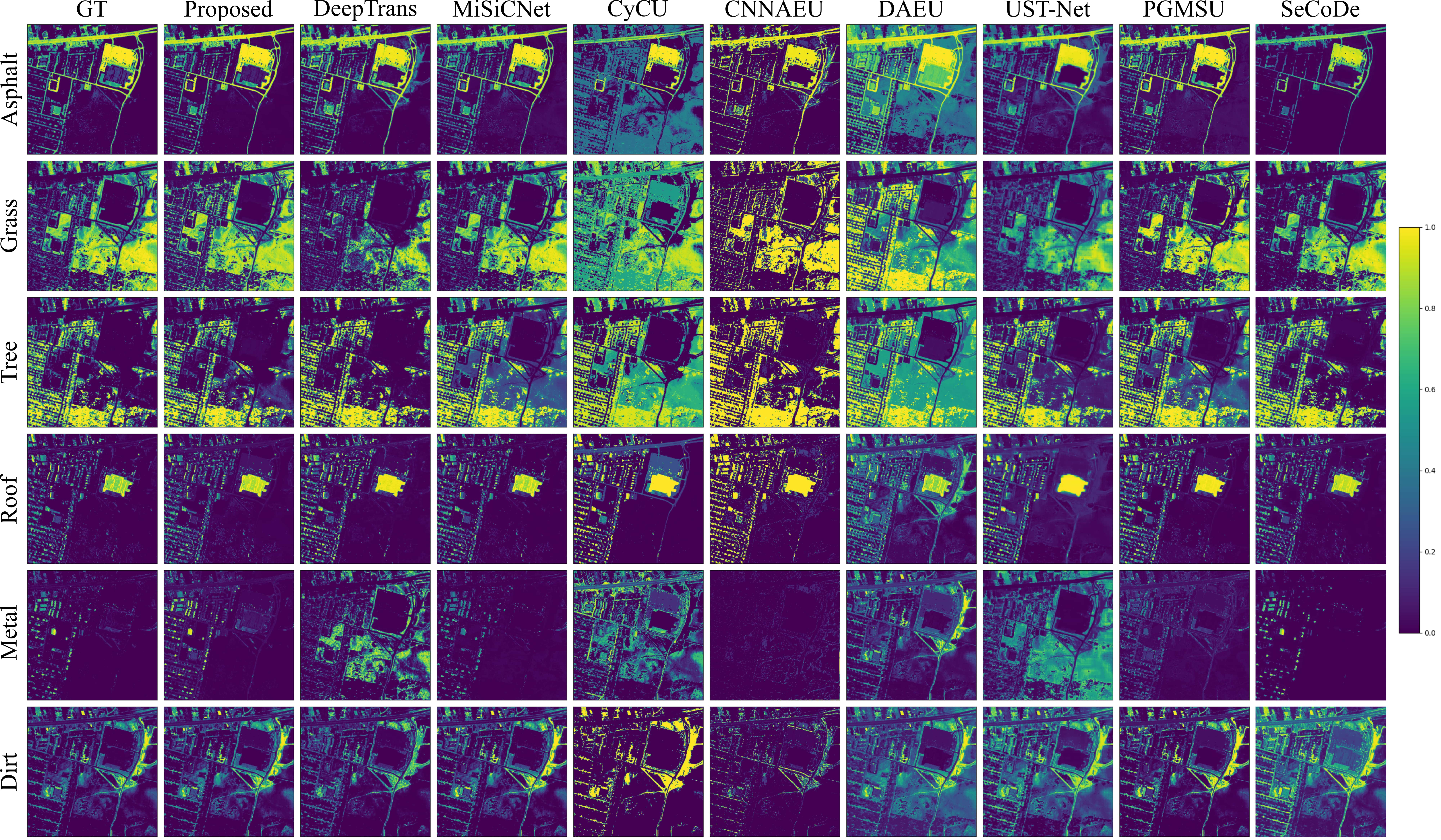}
    \caption{Abundance Maps Obtained by the Different Unmixing Algorithms for Urban Dataset}
    \label{urban_abd}
\end{figure*}
\begin{figure*}[!t]
    \centering
    \includegraphics[width=0.75\textwidth]{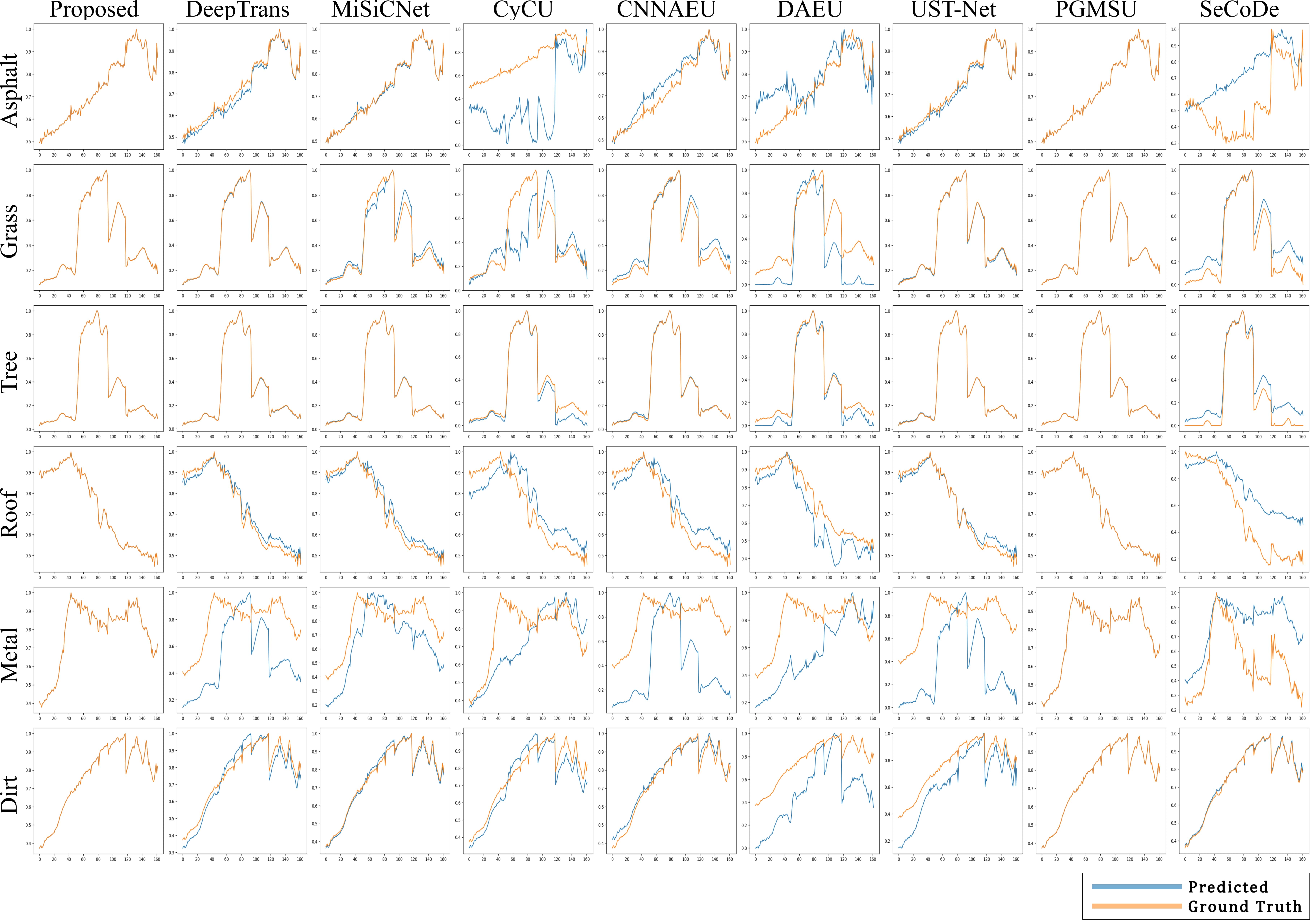}
    \caption{Endmember Spectral Signatures Obtained by the Different Unmixing Algorithms for Urban Dataset}
    \label{urban_sign}
\end{figure*}
\begin{table*}[!t]
\centering
\caption{Performance of Proposed Algorithm for Different Numbers of Endmembers for Urban Dataset}
\begin{tabular}{c|c|c|c}
\Xhline{2pt}
Error Metric & 4 Endmembers & 5 Endmembers & 6 Endmembers \\
\Xhline{2pt}
RMSE	&0.2500	&0.0962	&0.0727	\\
\hline
SAD	&0.1401	&0.0134	&0.0000\\
\Xhline{2pt}
\end{tabular}
\label{table: no_endmember}
\end{table*}

\acrlong{sp} is used to fuse the different EEAs to obtain a better spectral signature. For this fusion process, the endmember signatures extracted by N-FINDR, VCA, and ATGP have been used. When comparing the individual endmember signatures obtained from each EEA with the endmember signatures that the \acrlong{sp} predicted, the effectiveness of the process can be seen and a comparison for the Jasper-Ridge dataset is given in Figure \ref{eea_comparison}.   

In the Jasper-Ridge dataset, all EEAs were able to capture the tree endmember while only VCA and N-FNDR were able to capture the water endmember. Soil endmember was captured by both VCA and ATGP EEAs. However, none of these EEAs were able to capture the road endmember correctly as illustrated in Figure \ref{eea_comparison}.


As discussed in Section \ref{sp_architecture}, the model leverages an ensemble of endmembers derived from various \acrshort{eea}s as the initial basis for identifying the true endmembers. Figure \ref{eea_comparison} illustrates a comparison between the initial ensemble and the final endmembers for the Jasper-Ridge dataset. While none of the \acrshort{eea}s were able to perfectly capture the endmembers, the proposed algorithm succeeded in identifying highly accurate endmembers. Notably, this figure demonstrates that although each \acrshort{eea} managed to accurately identify certain endmembers, no single algorithm consistently outperformed the others across all signatures. This highlights the necessity of not relying on a single initialization method. The proposed algorithm was able to choose the best starting point for each endmember and further refine it to achieve highly accurate endmember signatures, surpassing what could be achieved by merely selecting or linearly combining the initial endmembers.



This is a good example to showcase the synergetic incorporation of the information of the ensemble to predict accurate endmemebers by the \acrshort{sp}. In otherwords even if the \acrshort{eea}s perform poorly the \acrshort{sp} is able to use the collective information provided by the ensemble to predict accurate endmember signatures.

In addition to the aforementioned performance metrics, another key aspect of the proposed algorithm is its performance under spectral variability. The article \cite{spectral_variability_jasper} discusses spectral variability in the Jasper-Ridge dataset. Table \ref{table: comparision_of_algorithms_jasper} shows that our algorithm performs well even with datasets that exhibit spectral variability.

\begin{figure*}[!t]
    \centering
    \includegraphics[width=0.75\textwidth]{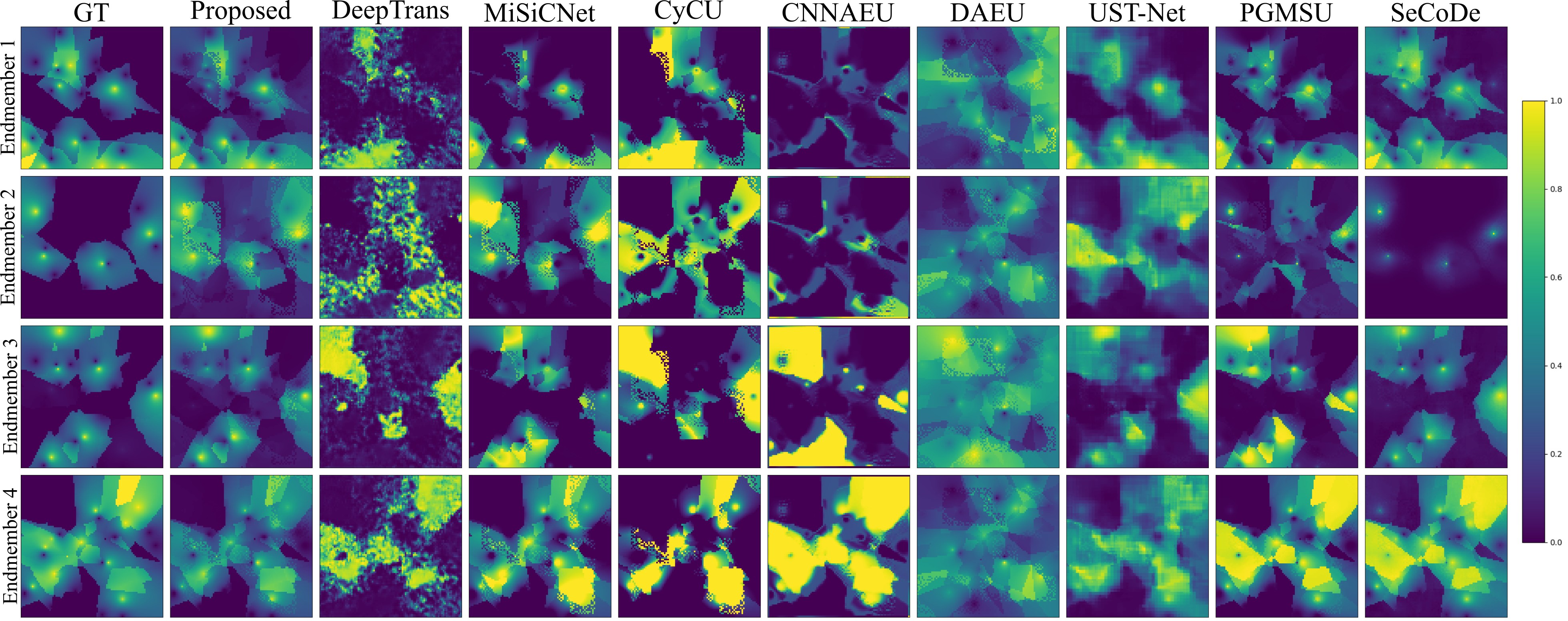}
    \caption{Abundance Maps Obtained by the Different Unmixing Algorithms for Synthetic Dataset}
    \label{synthetic_abd}
\end{figure*}
\begin{figure*}[!t]
    \centering
    \includegraphics[width=0.75\textwidth]{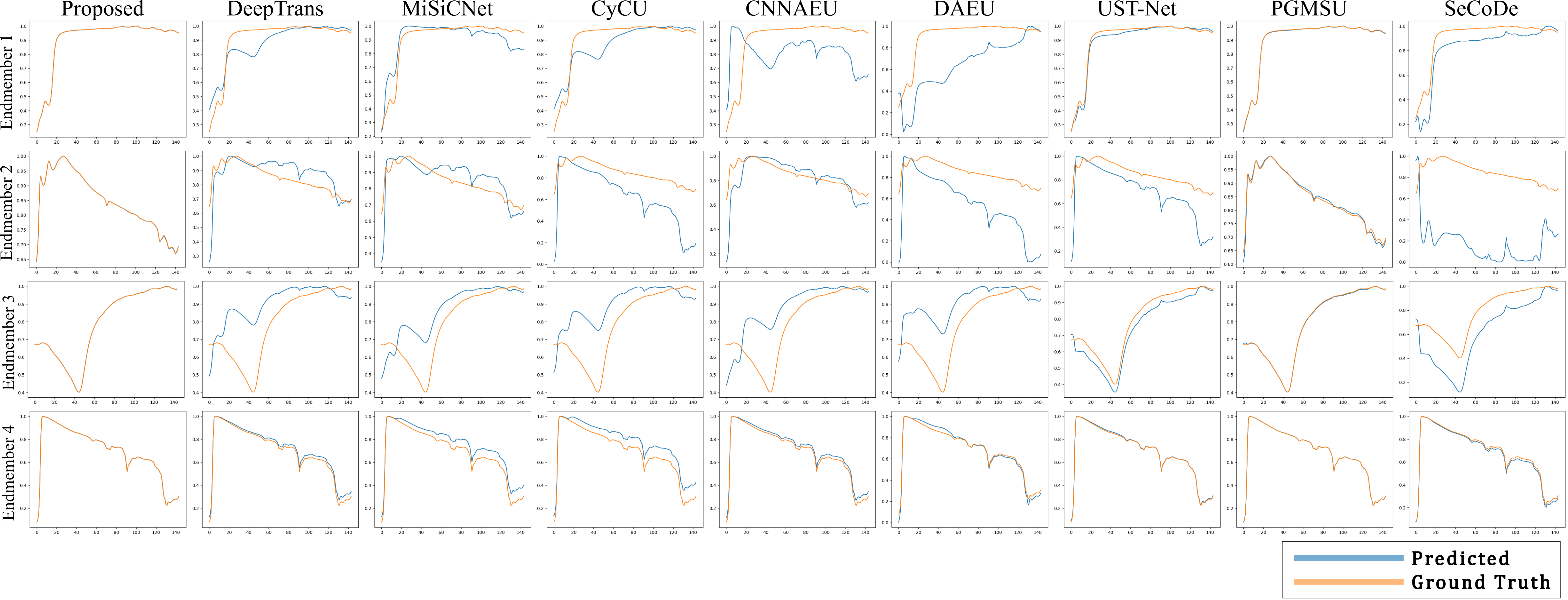}
    \caption{Endmember Spectral Signatures Obtained by the Different Unmixing Algorithms for Synthetic Dataset}
    \label{synthetic_sign}
\end{figure*}
\begin{table*}[!t]
\centering
\caption{Unmixing Performance Comparison for Different \acrfull{hs} Unmixing Algorithms for Synthetic Dataset.}
\footnotesize

\begin{tabular}{c |c| c| c| c| c| c| c| c| c| c}
\Xhline{2pt}
Error & Endmember & Proposed & DeepTrans & MiSiCNet & CyCU & CNNAEU & DAEU & USTNet & PGMSU & SeCoDe\\

\Xhline{2pt}
\multirow{5}{4em}{RMSE} & Endmember 1 
&0.0790\textsuperscript{2}	&0.2574	&0.2154	&0.2782	&0.3420	&0.3240	&0.0824\textsuperscript{3}	&0.1472	&\textbf{0.0730\textsuperscript{1}} \\
&Endmember 2 
&\textbf{0.1777\textsuperscript{1}} &0.4363 &0.2419	&0.2496	&0.3209	&0.2172\textsuperscript{3}	&0.2608	&0.2056\textsuperscript{2}	&0.2350 \\
& Endmember 3 
&\textbf{0.0495\textsuperscript{1}} &0.3269	&0.1651	&0.3884	&0.3329	&0.5147	&0.1072\textsuperscript{3}	&0.1918	&0.0780\textsuperscript{2} \\
& Endmember 4 
&0.1464\textsuperscript{2}	&0.2111	&0.1628\textsuperscript{3} &0.2838	&0.2320	&0.3911	&0.2114	&\textbf{0.1344\textsuperscript{1}}	&0.1831 \\
&Average	
&\textbf{0.1242\textsuperscript{1}}	&0.3194	&0.1992	&0.3046	&0.3101	&0.3775	&0.1809	&0.1723\textsuperscript{3}	&0.1582\textsuperscript{2} \\
[1ex] 
\hline

\multirow{5}{4em}{SAD} & Endmember 1 
&\textbf{0.0000\textsuperscript{1}}	&0.0867	&0.0858	&0.0921	&0.2465	&0.2547	&0.0190\textsuperscript{3}	&0.0020\textsuperscript{2} &0.0833\\
& Endmember 2
&\textbf{0.0000\textsuperscript{1}}	&0.1088	&0.0881\textsuperscript{3}	&0.2793	&0.1356	&0.3643	&0.2021	&0.0100\textsuperscript{2}	&0.8201\\
& Endmember 3 
&\textbf{0.0000\textsuperscript{1}}	&0.1738	&0.1334	&0.1629	&0.1655	&0.1625	&0.0358\textsuperscript{3}	&0.0020\textsuperscript{2}	&0.1825\\
& Endmember 4 
&\textbf{0.0000\textsuperscript{1}}	&0.0252	&0.0527	&0.0653	&0.0228	&0.0324	&0.0049 \textsuperscript{3}	&\textbf{0.0000\textsuperscript{1}}	&0.0143 \\
& Average 
&\textbf{0.0000\textsuperscript{1}}	&0.0986	&0.0900	&0.1499	&0.1426	&0.2035	&0.0654\textsuperscript{3}	&0.0035\textsuperscript{2}	&0.2751 \\
[1ex] 
\Xhline{2pt}
\end{tabular}

\label{table: comparision_of_algorithms_synthetic}
\end{table*}

\begin{table*}[!t]
\centering
\caption{Perfomance of Proposed Method Without the \acrlong{an} for Samson Dataset}
\begin{tabular}{c|c|c|c|c|c}
\Xhline{2pt}
Configuration &Error Metric & Soil & Tree & Water & Average \\
\Xhline{2pt}
\multirow{2}{*}{without \acrshort{an} }&RMSE	&0.0656	&0.0448	&0.0293	&0.0489\\
&SAD	&0.0136	&0.0290	&0.0328	&0.0251\\
\Xhline{2pt}
\multirow{2}{*}{with \acrshort{an} }&RMSE	&0.0442	&0.0266	&0.0258	&0.0333\\
&SAD	&0.0117	&0.0309	&0.0326	&0.0250\\
\Xhline{2pt}
\end{tabular}
\label{table: without_an}
\end{table*}
\begin{table*}[!t]
\centering
\caption{Performance of Proposed Method Under Different Neighborhood Configurations for Samson Dataset}
\footnotesize

\begin{tabular}{c |c| c| c| c| c| c| c| c| c| c}
\Xhline{2pt}
\multicolumn{2}{c|}{}& \multicolumn{3}{c|}{Doughnut} & \multicolumn{3}{c|}{Circle} & \multicolumn{3}{c}{Normal}\\
\hline
Error & Endmember & level 3 & level 4 & level 5 & level 3 & level 4 & level 5 & level 3 & level 4 & level 5\\
\Xhline{2pt}
\multirow{4}{4em}{RMSE} & Soil 
&0.0516	&0.0703	&0.0754	&0.0740	&\textbf{0.0443}	&0.0553	&0.1217	&0.0932	&0.0975\\
&Tree 
&0.0306	&0.0409	&0.0503	&0.0406	&\textbf{0.0267}	&0.0328	&0.1267	&0.0844	&0.0799\\
& Water 
&0.0267	&0.0415	&0.0376	&0.0438	&\textbf{0.0258}	&0.0308	&0.0367	&0.0268	&0.0389\\
&Average	
&0.0379	&0.0528	&0.0566	&0.0549	&\textbf{0.0334}	&0.0412	&0.1036	&0.0742	&0.0762\\
[1ex] 
\hline

\multirow{4}{4em}{SAD} & Soil 
&0.0122	&\textbf{0.0103}	&0.0119	&0.0170	&0.0117	&0.0142	&0.0157	&0.0141	&0.0335\\
& Tree
&0.0326	&0.0339	&0.0326	&0.0368	&0.0309	&0.0333	&0.0311	&\textbf{0.0294}	&0.0298\\
& Water 
&0.0331	&0.0363	&0.0310	&\textbf{0.0301}	&0.0326	&0.0302	&0.0312	&0.0305	&0.0387\\
& Average 
&0.0260	&0.0268	&0.0252	&0.0280	&\textbf{0.0251}	&0.0259	&0.0260	&0.0246	&0.0340\\
[1ex] 
\Xhline{2pt}
\end{tabular}

\label{table: an_configuration}
\end{table*}


\begin{table*}[!t]
\centering
\caption{Performance of Proposed Algorithm for Different Stages of Training for Jasper Dataset}
\footnotesize

\begin{tabular}{c |c| c| c| c}
\Xhline{2pt}
Error & Endmember & Stage 1 Only & Stage 2 Only & Stage 1 and 2 \\
\hline 
\multirow{4}{4em}{RMSE} & Tree 
&0.1833 &0.3064 &\textbf{0.0822}\\
& Water 
&0.4923 &0.3755 &\textbf{0.0807}\\
& Soil 
&0.3829 &0.2749 &\textbf{0.0776}\\
& Road 
&0.2277 &0.2195 &\textbf{0.1048}\\
&Average	
&0.3444 &0.2995 &\textbf{0.0854}\\
[1ex] 
\hline

\multirow{4}{4em}{SAD} & Tree 
&0.1590 &0.9302 &\textbf{0.1222}\\
& Water 
&0.7734 &1.6615 &\textbf{0.1355}\\
& Soil 
&0.2864 &0.1901 &\textbf{0.0427}\\
& Road 
&0.0620 &2.2598 &\textbf{0.0421}\\
&Average	
&0.3202 &1.2604 &\textbf{0.0870}\\
[1ex] 
\Xhline{2pt}
\end{tabular}

\label{table: different_training_stages}
\end{table*}

\subsection{Experiment with Urban Dataset}
\subsubsection{Perfomance Comparison with the State-Of-The-Art Algorithms}

When considering the Urban dataset, the ground truth abundance maps and predicted abundance maps for each algorithm are shown in Figure \ref{urban_abd}. The comparison of the predicted signatures with the ground truth for each algorithm is given in Figure \ref{urban_sign}. The error metric for each of the algorithms is given in table \ref{table: comparision_of_algorithms_urban} and the time taken for training and the prediction are tabulated in \ref{table: time_cost}.

If we consider the \acrshort{rmse} values it is clear that the proposed algorithm has the best average result. Considering the \acrshort{rmse} of the individual endmembers the proposed algorithm shows the best values for the grass, metal and dirt endmembers while showing comparable results for the other endmembers. 
If we consider the \acrshort{sad} values the proposed algorithm was able to extract all the endmember signatures perfectly. In addition to this PGMSU algorithm was also able to identify all the endmember signatures perfectly. 


\subsubsection{Performance Comparison with Different Number of Endmembers}
To analyze the effect of the number of endmembers on the proposed algorithm, the Urban dataset with ground truth for 4, 5, and 6 endmembers was used. The results, as shown in the table \ref{table: no_endmember}, indicate that the proposed algorithm achieves better results as the number of endmembers increases. This improvement is expected since omitting certain endmembers can lead to data that does not accurately represent reality, thereby hindering performance. Conversely, when the number of endmembers more accurately reflects reality, the proposed algorithm's performance improves.

\subsection{Experiment with Synthetic Dataset}

\subsubsection{Perfomance Comparison with the State-Of-The-Art Algorithms}
For the Synthetic dataset, the ground truth and predicted abundance maps for each algorithm are shown in Figure \ref{synthetic_abd}. The comparison of the predicted signatures with the ground truth for each algorithm is presented in Figure \ref{synthetic_sign}. The error metrics for each algorithm are detailed in Table \ref{table: comparision_of_algorithms_synthetic}, and the training and prediction times are shown in Table \ref{table: time_cost}.

Regarding \acrshort{rmse} values for abundances, the proposed algorithm shows the best overall performance. Specifically, it achieves the lowest \acrshort{rmse} values for the second and third endmembers and the second-best results for the other two endmembers.

In terms of \acrshort{sad} values, the proposed algorithm has identified the endmembers perfectly. This indicates that, under ideal conditions, the algorithm performs exceptionally well.

\subsection{Performance of Different Network Configurations}

\subsubsection{Different Neighborhood Settings}

The experiment was done with three different neighborhood shapes and three different size levels. They are doughnuts, circles, and random shapes with a normal distribution. It was found that the circle shape with level 4 gives the best performance for the Samson dataset. A comparison between the shapes and different padding levels is given in Table \ref{table: an_configuration}. It can be seen that the other two shapes also give competitive results. 
 


\subsubsection{Influence of Training Stages}

In this section the requirement of two training stages for the training of the \acrshort{ap} and \acrshort{sp} is analyzed. For this purpose, the \acrshort{ap} and \acrshort{sp} were trained in the following configurations for the Jasper Dataset.
    \begin{enumerate}
        \item Only stage 1 training.
        \item Only stage 2 training.
        \item Both stage 1 and stage 2 training.
    \end{enumerate}

The results for the above configurations is summarized in Table \ref{table: different_training_stages}. These results underscore the necessity of employing two training stages, as the most significant improvements are observed when both stages are utilized. Upon closer examination, it becomes evident that using only training stage 2 leads to poor \acrshort{sad} values, indicating erroneous selection of endmember spectra. This highlight the importance of training stage 1 in guiding the \acrshort{ap} initially, as discussed in Section \ref{training process}. In such cases, the model may converge to a local minimum, as suggested by the suboptimal abundance values, failing to identify the correct endmember signatures.

Conversely, when only stage 1 of training is employed, there is a noticeable improvement in \acrshort{sad} values compared to using only stage 2. However, the outcomes are still suboptimal since the \acrshort{sp} only generates linear combinations of the provided ensemble, resulting in suboptimal predictions for signatures. This naturally leads to suboptimal abundance values.

Ultimately, when stage 2 follows stage 1 in training, substantial enhancements are observed in both endmember signature \acrshort{sad} values and abundance \acrshort{rmse} values. Here, stage 1 directs the \acrshort{ap} appropriately by utilizing the constrained \acrshort{sp}. Subsequently, stage 2 empowers the \acrshort{sp} to update the ensemble itself, moving beyond linear combinations. This dual approach allows both \acrshort{ap} and \acrshort{sp} to converge effectively on accurate signatures and abundances, resulting in optimal outcomes.

It should be noted that for certain datasets training stage 2 may not be necessary since training stage 1 alone may lead to the optimal result. An example of this is the Synthetic Dataset used in experiments.

\section{Conclusion}
\label{conclusion}
In this paper, we propose a novel transformer-based deep learning algorithm for \acrshort{hs} unmixing that is capable of fusing endmember signatures extracted by different endmember extraction algorithms to predict better endmember signatures. In addition to that, the proposed method can take spatial context into account in the training process which leads to higher accuracy in abundance prediction. This algorithm has been tested for three popular \acrshort{hs} datasets and one synthetic dataset, and the results have been compared with the state-of-the-art algorithms. Using the obtained results, we demonstrated the ability to leverage the attention mechanism of the transformer to not only inject the spatial context for better abundance map generation but also synergetically fuse endmembers obtained from endmember extraction algorithms resulting in better performance. Therefore, we assume that this work will promote more research along this direction, of utilizing transformer-based endmember fusion for remote sensing.


\printbibliography


\end{document}